\documentclass[%
aip,
amsmath,amssymb,
preprint,
]{revtex4-1}
\usepackage[letterpaper, portrait, margin=1in]{geometry}
\usepackage{graphicx}
\usepackage[version=4]{mhchem}
\usepackage{xurl}
\usepackage{bm}
\usepackage{bbm}
\usepackage{hyperref}
\hypersetup{colorlinks=true,urlcolor=blue,citecolor=blue,linkcolor=blue}
\usepackage{jabbrv}

\usepackage{wrapfig}
\def\bx{{\bf x}}
\def \bp{{\bf p}}
\def \bP{{\bf P}}
\def \bff{{\bf f}}
\def \bs{{\bf s}}

\begin{document}

\title{Markov State Models for Tracking Reaction Dynamics on Catalytic Nanoparticles}

\author{Caitlin A. McCandler}
\address{Initiative for Computational Catalysis, Flatiron Institute, NY 10010, USA}
\author{Chatipat Lorpaiboon}
\address{Initiative for Computational Catalysis, Flatiron Institute, NY 10010, USA}
\author{Timothy C. Berkelbach}
\address{Initiative for Computational Catalysis, Flatiron Institute, NY 10010, USA}
\address{Department of Chemistry, Columbia University, NY 10027, USA}
\author{Jutta Rogal}
\address{Initiative for Computational Catalysis, Flatiron Institute, NY 10010, USA}

\date{\today}

\begin{abstract}
Markov state models (MSMs) are a powerful tool to analyze and coarse-grain complex dynamical data into interpretable kinetic processes. This capability is particularly important in heterogeneous catalysis, where a medley of reactants and intermediates interact on surfaces that might simultaneously experience structural fluctuations. 
For these very complex systems, standard transition state theory (TST) approaches are no longer appropriate, motivating alternative approaches that can retain dynamical complexity while providing physical insight. With machine learned interatomic potentials being more and more ubiquitous, directly simulating complex catalytic systems with molecular dynamics (MD) is becoming increasingly feasible. Extending MSMs to dynamically coarse grain MD simulation data of catalytic processes, we analyze hydrogen dynamics on rhodium catalysts with slab and nanoparticle geometries over a range of hydrogen surface concentrations. Somewhat counterintuitively,  nanoparticle features, such as corners and edges, effectively slow down the association/dissociation process, and the cooperative behavior of hydrogen-hydrogen interactions leads to a non-monotonic concentration dependence of the rates, which would not be predicted with standard TST. 
\end{abstract}

\maketitle

\section{Introduction}
Uncovering the mechanisms that drive the selectivity and efficiency of catalytic reactions is essential when aiming to improve known catalysts and propose new catalyst materials. 
The catalytic activity of a material depends on its composition and structure, as well as operating conditions like temperature and pressure. Typical theoretical approaches to determining reaction mechanisms and rates with \emph{ab initio} accuracy identify the so-called rate limiting step, calculate the energy barriers of the reactive events, and estimate the corresponding rate constants within harmonic transition state theory (TST).~\cite{vineyard_frequency_1957}
For example, when analyzing a dissociative adsorption reaction of a small molecule on a catalytic surface, great care is taken to find the optimal binding orientation, the lowest energy transition state, and the vibrational frequencies of reactants and intermediates, such that one can calculate accurate adsorption and dissociation energies.~\cite{ledentu_ab_1998,eichler_ab-initio_1998,mccormack_mechanisms_2005,feibelman_orientation_1991,eichler_quantum_1996} However, these calculations are often of a single adsorbate on an otherwise bare, defect-free surface. 

As a step beyond this static picture and to capture the dynamical evolution and statistical interplay of various processes, kinetic Monte Carlo (KMC) models have been successfully applied to catalytic systems.~\cite{sabbe_first-principles_2012,reuter_ab_2016,latz_three-dimensional_2012} This type of first-principles-based microkinetic modeling usually builds on a predefined event table of possible  processes for which the corresponding rate constants are again determined by combining electronic structure calculations with hTST.
However, under operating conditions, the structure of the catalyst is under constant change, continuously creating and destroying active sites on timescales that might be comparable to the catalytic reaction itself.~\cite{tao_situ_2011,bonati_role_2023,zhang_non-equilibrium_2025} Additionally, the concentration and number of chemical species interacting on the catalyst surface can widely fluctuate during operation.
Identifying all possible atomistic processes and their corresponding energy barriers for such complex scenarios by hand is tedious at best and impossible in most cases.
Recently, approaches that automate the exploration of new states and identify the corresponding processes and rate constants together with an iterative refinement of a machine-learned interatomic potential (MLIP) for fast energy evaluations have tried to alleviate some of these issues.\cite{lai_automatic_2025,poths_ml-accelerated_2025,trushin_self-learning_2005,el-mellouhi_kinetic_2008,xu_adaptive_2008} 
However, the total number of atomic environments and elementary steps to be included quickly explodes combinatorially with system size, making these methods intractable for large complex systems. 
This is particularly true when both the dynamics of the catalytic reaction as well as of the restructuring of the catalyst material need to be considered simultaneously.

Here, we consider a different route, starting from the full dynamics of the entire system in the phase space and subsequently extracting the slow degrees of freedom from the molecular dynamics (MD) trajectories. 
Specifically, we apply Markov state models (MSMs)~\cite{noe_constructing_2009,prinz_probing_2011} to dynamically coarse grain the simulation data 
and retrieve interpretable insights on, for example, the reaction conditions that lead to high activity, those that lead to poisoning, and the role of surface features. 
A key advantage of this approach is that it does not require defining any reactive events a priori, and it allows one to naturally capture the complex interplay of various dynamical processes.
MSMs have been applied successfully in the biophysics community to determine, for example, the timescales of protein folding,\cite{husic_markov_2018,arbon_markov_2024} where the simplified picture of unique and well defined transition states on the potential energy surface is inadequate. 
Traditionally, MSMs have been used to model global changes of the entire system (e.g., folded vs.~unfolded) but 
have recently also been adopted to understand the kinetics of more local processes~\cite{xie_graph_2019,li_unveiling_2025}---an idea that extends nicely to the study of heterogeneous catalysis. Additionally, due to the rapid advancement of MLIPs, making them much more accurate, fast, and accessible, it is becoming possible to simulate heterogeneous catalysis over an extended timescale with sufficient sampling of reactive events. As the amount of simulation data in this space increases with the help of MLIPs, MSMs will allow us to draw relevant information from the dynamical trajectories, which may support catalyst design. 

To illustrate our approach, we examine hydrogen dissociation on rhodium surfaces and nanoparticles, Fig.~\ref{fig:MSMconstruction}A,
aiming to disentangle the complex effects of operating pressure and surface features on the reactivity of hydrogen on rhodium catalysts. The analysis of the MSMs for the different systems reveals that the rates of \ce{H2} dissociation and association are indeed different on nanoparticles as compared with simple slab facets. Notably, we find that nanoparticles maximize their catalytic activity by under-saturating the surface. In each system, the contributions to the reactivity can also be mapped to certain states, increasing our understanding of the effects that can improve activity.

\section{Markov State Models for Catalysis}
In molecular dynamics, the time evolution of a configuration $\bx$ is determined by the potential energy function, the system setup, and the numerical integrator for the equations of motion. In general, the dynamics of a molecular system in the phase space are non-linear but according to Koopman theory~\cite{brunton_modern_2022} there exists a coordinate transformation into some feature space $\bm\chi(\bx)$ in which the dynamics are generated by a linear operator. The advantage of such a linear operator is that the dynamics can now be fully characterized by analyzing its eigenvectors and eigenvalues. If the feature transformation is given by indicator functions into a discrete state space, $\chi_i(\bx) = \mathbbm{1}_i(\bx)$, the corresponding linear dynamics over a lag time $\tau$ are
\begin{equation}\label{eq:lineardyn}
\bp_{t+\tau} = \bP^{\top}(\tau)\bp_{t} \quad ,    
\end{equation}
where $\bp$ is the vector of state probabilities with elements $p_i = \mathbb{E}[\chi_i(\bx)] = \mathbb{E}[ \mathbbm{1}_i(\bx)]$, where $\mathbb{E}[ \dots ]$ denotes the expectation over the ensemble, and $\bP$ is the matrix of transition probabilities between states. In a system with separate timescales for different processes, the eigenvalue spectrum of $\bP$ will have a gap indicating the separation between slow and fast dynamical modes. 
When building MSMs from MD simulations to coarse grain the dynamics and analyze slow degrees of freedom, the transition matrix $\bP$ is estimated from the trajectory data. 
For a more detailed  derivation of  MSMs, we point the reader to some excellent reviews on this topic.~\cite{husic_markov_2018,prinz_markov_2011,wang_constructing_2018} 

 The central element when building MSMs is the identification of a suitable mapping  $\bm\chi(\bx)$ which is largely system dependent.
To apply MSMs to catalysis, the main conceptual extension is the idea that the global system's dynamics can be inferred from the aggregation of the dynamics of the individual reactants. Instead of following global changes in the entire system, only changes in the local environments of each reactant are tracked which can be realized by choosing an atom-centered local featurization, correspondingly.~\cite{li_unveiling_2025,xie_graph_2019}  
Such a local state space representation allows us to focus on the dynamical processes most relevant for reactive events while
 capturing the remaining degrees of freedom implicitly.
Looking at diffusion as an example, this means that an atom moving between identical adsorption sites on a surface remains in the same state. However, if the local environment around an atom changes---e.g., due to structural features such as edges or defects, proximity to other adsorbates, or reactive events---the dynamics of moving between these states is explicitly captured by the model.  

\begin{figure}
    \centering
    \includegraphics[width=\linewidth]{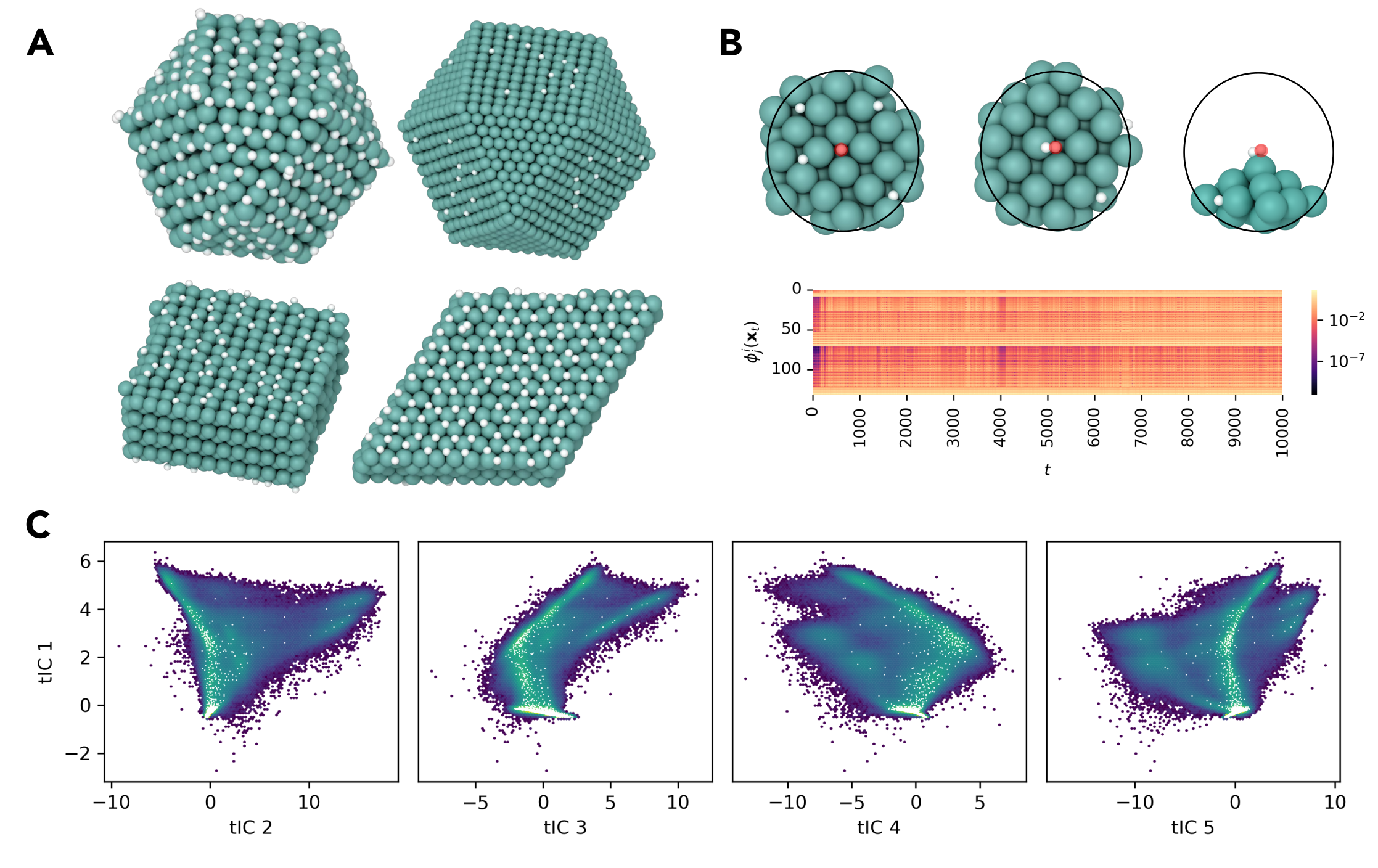}
    \caption{Markov State Models for hydrogen dissociation/association on rhodium. A) Simulation data is generated for hydrogen atoms interacting with rhodium surfaces with four different geometries (2 nm nanoparticle, 5 nm nanoparticle, (100) slab, (111) slab) and various H concentrations. 
    B) The local environment around each hydrogen atom $i$ is represented by a feature vector $\bm \phi^i(\bx_t)$. The feature values, $\phi_j^i(\bx_t),$ change as a function of time reflecting changes in local coordination and composition. 
    C) Density plot of TICA projection of the high dimensional ($d_\phi = 131$) feature space into the five dynamically most relevant dimensions (tIC 1-5); Kmeans++ clustering is performed within this lower dimensional space to discretize the states, white points indicate cluster centers.}
    \label{fig:MSMconstruction}
\end{figure}

Given a trajectory $\bx^T = \{\bx_1,\dots,\bx_T\}$ of length $T$ in the full configuration space, a set of features $\bm \phi$ representing the local environment around each reactant, Fig.~\ref{fig:MSMconstruction}B, is calculated 
\begin{equation} \label{eq:features}
\bx_t \in \mathbb{R}^{3N} \rightarrow \bm\phi^i(\bx_t)  \in \mathbb{R}^{d_\phi} \quad ,
\end{equation}
resulting in individual trajectories for each atom $i$ included in the analysis. For the hydrogen-rhodium system, local features are only computed for the hydrogen atoms.
The representation of the local environments should be smooth with respect to the interaction cutoff distance, and should be expressive enough to distinguish between states. Here, a subset of the functions used to construct the MLIP is employed as they satisfy both requirements (see SI Sec.~\ref{subsec:msmfitting}), providing a hydrogen atom-centered local structure representation, $\bm \phi^i(\bx)$, with $d_\phi = 131$ features corresponding to H and Rh 2-body, 3-body, and 4-body interactions with a cutoff distance of 7~\AA. 
The variation of the feature values for a single hydrogen atom along a piece of a trajectory is schematically depicted in Fig.~\ref{fig:MSMconstruction}B. 

The subsequent construction of the MSM follows the same procedure as is employed for MSMs based on global features. First, the generally high dimensional feature space, $\bm\phi^i(\bx)$, is reduced to a lower dimensional space, $\bff^i \in \mathbb{R}^{d_f}$ usually with $d_f \leq 8$, while trying to retain as much dynamical information as possible. Here, time-lagged independent component analysis (TICA)\cite{molgedey_separation_1994,perez-hernandez_identification_2013} is used to reduce the $d_\phi = 131$ local representation features into only $d_f = 5$ TICA dimensions, separating the states by their dynamical distance. Reducing the dimensions is necessary to cluster the trajectory data into discrete states, but reducing to too few dimensions will remove important dynamical information. The optimal number of TICA dimensions varies for each system and needs to be carefully tested. 
Kmeans++ clustering~\cite{arthur_k-means_2007} is performed in TICA space with $d_s=1200$ cluster centers (shown as white points in Fig.~\ref{fig:MSMconstruction}C) representing the discrete microstates of the MSM. 
When analyzing the trajectory data, the local environment around each H atom, $\bm \phi^i(\bx_t)$, is mapped to a microstate $s_j(\bx_t) =  \mathbbm{1}_j(\bx_t)$
\begin{equation}\label{eq:statemap}
\bm \phi^i(\bx_t) \in \mathbb{R}^{d_\phi}\rightarrow \bff_t \in \mathbb{R}^{d_f} \rightarrow \bs_t \in \mathbb{R}^{d_s} \quad ,
\end{equation}
and the resulting discrete state trajectory $\bs^T=\{\bs_1,\dots,\bs_T\}$ is used to estimate the transition matrix $\bP(\tau)$ in Eq.~\eqref{eq:lineardyn} for a given lag time $\tau$.
The eigenvectors of the transition matrix correspond to the coarse-grained dynamical modes of the system, and the eigenvalues, $\lambda$, provide the relaxation timescales of the dynamical processes for a given $\tau$
\begin{equation}
\label{eq:timescale}
t(\tau) = -\frac{\tau}{\ln|\lambda(\tau)|} \quad .
\end{equation}
A schematic illustrating each of these steps in the MSM construction is included in Fig.~\ref{fig:MSMconstruction}. For each simulation setup (four different geometries and various hydrogen concentration), a separate MSM was fitted because the timescales of the processes in each system were so diverse. As the results can be sensitive to the details of the MSM construction,~\cite{arbon_markov_2024} the hyperparameters (e.g. TICA dimensions, cluster centers, and lag times) were carefully tested to yield robust outcomes over all the MSM fittings (see SI Sec.~\ref{subsec:msmfitting}). The MSMbuilder package was used to create the MSMs.~\cite{harrigan_msmbuilder_2017}

\section{Dynamical Analysis of the Rh-H system}
\subsection{Dynamically Distinct Hydrogen States on Rhodium} 
The analysis of the fitted MSMs provides information regarding the most relevant dynamical processes in the investigated systems. In this section, the derivation of the computed quantities is described for a representative example Rh-H system: a large nanoparticle (5 nm) with about 50\% hydrogen coverage at $T=450$~K. All simulations were performed using an Atomic Cluster Expansion (ACE)\cite{drautz_atomic_2019} MLIP fitted to \textit{ab initio} data obtained with DFT using the PBE functional (see SI Sec.~\ref{subsec:mlip}), and the MD simulations to generate trajectory data were performed in LAMMPS~\cite{thompson_lammps_2022} with a Langevin thermostat (see SI Sec.~\ref{subsec:simsetting}).

\begin{figure}
    \centering
    \includegraphics[width=\linewidth]{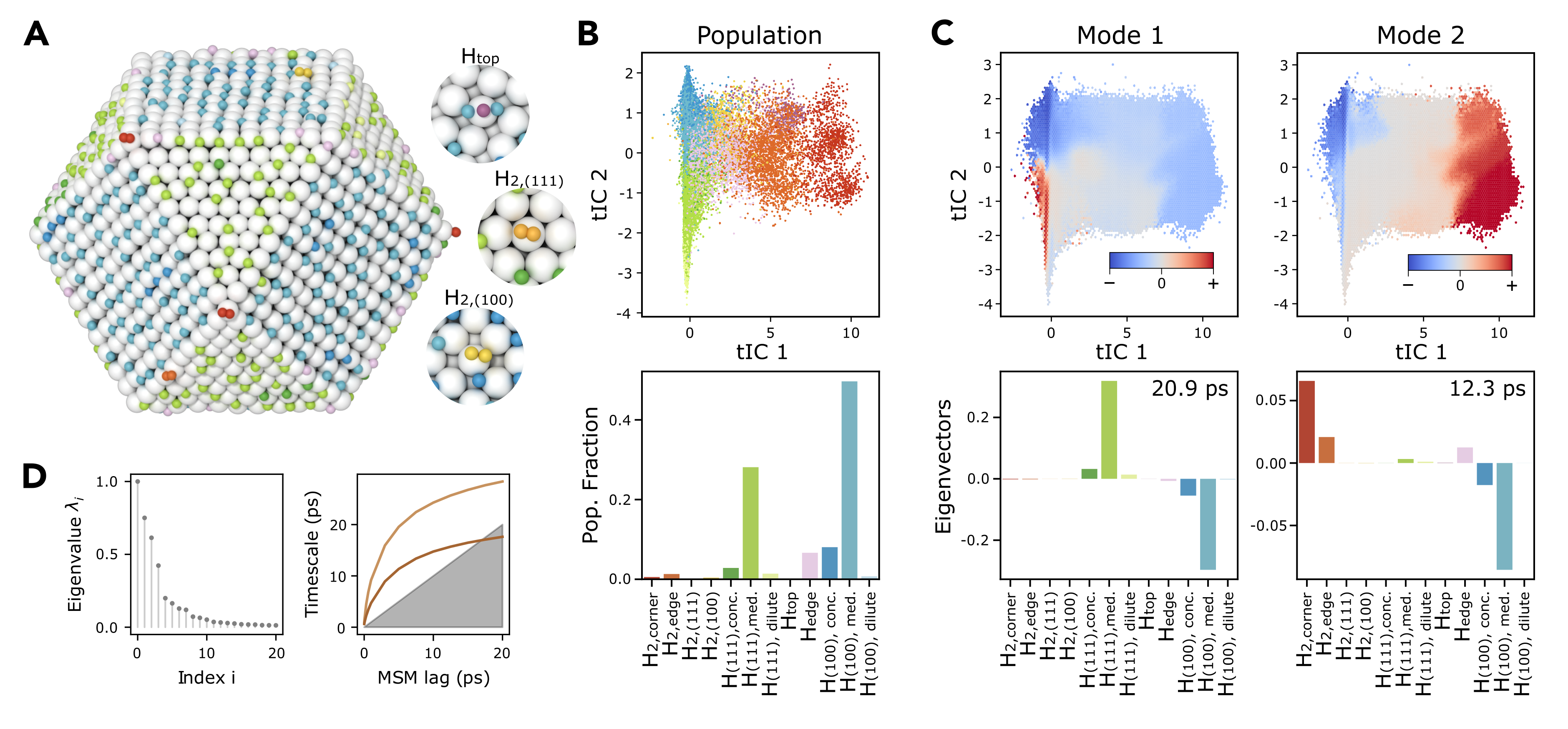}
    \caption{The slowest dynamical processes of hydrogen atoms interacting on a large rhodium nanoparticle. The surface coverage of hydrogen in this example is about 50\%. A) Simulation snapshot where hydrogen atoms (small spheres) are colored by their geometric state label and rhodium atoms are represented white, large spheres. \ce{H2} on corners/edges/facets are red/orange/yellow and H on (100)/(111)/edges/top sites are blue/green/light purple/dark purple. The local concentration of H on surfaces is also indicated by dark/light being concentrated/dilute. B) Labeled data projected into TICA space (sampled every ps, top graph) and a bar plot of the equilibrium distribution obtained from the values of the first eigenvector and integrated over the corresponding microstates. C) Projected flux of the two slowest dynamical modes and the states that contribute to these modes. The slowest process is the diffusion from a (111) facet to a (100) facet. The second slowest process is the association/dissociation reaction of \ce{H}$\leftrightarrow$$\frac{1}{2}$\ce{H2}. D) Spectral plot of the eigenvalues of the transition matrix and the implied timescale convergence of the two slowest processes.}
    \label{fig:flux}
\end{figure}

\subsubsection{Geometric State Assignment}
To correlate the dynamical modes of the MSMs with structural characteristics of the Rh-H systems, each hydrogen atom was categorized based on geometric features as belonging to one of the following twelve states: \ce{H2} dimers on corners/edges/(100) facets/(111) facets,  \ce{H} atoms on (100)/(111) facets with many/some/few neighbors, H atoms at nanoparticle edges, or H atoms sitting atop Rh atoms. Figure~\ref{fig:flux}A shows a simulation snapshot where the hydrogen atoms are colored according to these labels. These structural state assignments are indeed correlated with dynamically distinct states, as they are well separated when projected into TICA space (top graph in Fig.~\ref{fig:flux}B). Labeling the conformations according to these structural macrostates is only for the purpose of interpretability. Any approximation of the higher dimensional state space ($d_s = 1200$ in our case) always reduces the quality of the MSM.~\cite{prinz_probing_2011}  

Interestingly, a structural motif that has not been discussed before (to the best of our knowledge) emerged as a dynamically distinct state: $\text{H}_{\text{top}}$. In this state, the hydrogen sits on top of a rhodium atom with two close neighbor hydrogen atoms which act as a trap for the hydrogen in the otherwise unfavorable top site. This $\text{H}_{\text{top}}$ state forms on both facet types ((111)/(100)) as well as on nanoparticle edges, but was never observed on a nanoparticle corner site. 
To ensure that this is not an artifact of the employed MLIP, additional training data was included in the MLIP fit (see SI Sec.~\ref{subsec:mlip}). 

\subsubsection{Dynamical Modes and Timescales}
Analysis of the MSM eigenvectors and eigenvalues showed that four distinct dynamical modes account for the majority of the slow dynamics in the Rh-H systems: adsorption and desorption, association and dissociation, diffusion between (111) and (100) facets, and exchange between regions of concentrated and dilute hydrogen surface coverage. To focus on dynamical processes of hydrogen adsorbed on the surface, all adsorption/desorption processes were removed from the trajectory data, assuming an equilibrium coverage for a given H concentration  (see SI Sec.~\ref{subsec:msmfitting}). 

The spectral plot of the first 20 eigenvalues, $\lambda$, of the transition matrix (left graph in Fig. \ref{fig:flux}D) shows a gap after the fourth eigenvalue indicating that there are three dominant slow modes after removing the adsorption mode. 
The largest eigenvalue is always equal to $\lambda_0 = 1$ and corresponds to an infinite equilibration timescale. The respective left eigenvector yields the equilibrium distribution in the space of discrete microstates $\bs \in \mathbb{R}^{d_s}$. By grouping configurations assigned to a microstate according to their structural labels, we obtain a more intuitive representation of the equilibrium distribution shown in the bottom graph of Fig.~\ref{fig:flux}B. 
All other eigenvalues and eigenvectors provide the relaxation timescales (through Eq.~\eqref{eq:timescale}) and processes (flux between microstates) of the dynamical modes. 
Ideally, the timescales should be independent of the chosen lag time which is usually assessed by plotting the timescales as a function of $\tau$ (right graph in Fig.~\ref{fig:flux}D).
A projection of the flux between microstates of the two slowest processes into the first and second TICA dimension is shown in the top graphs of Fig.~\ref{fig:flux}C. Again, a more intuitive picture can be obtained by evaluating the flux between structural state classifications integrated over the microstates, as depicted in the bottom graphs of Fig.~\ref{fig:flux}C.
The slowest mode in this example is clearly the exchange of H atoms between (100) and (111) facets since the majority of the positive/negative flux in the second eigenvector corresponds to states that have been labeled as H$_\text{(111)}$/H$_\text{(100)}$.
The second slowest mode is associated with dissociation/association of \ce{H2} on corners and edges and H atoms on (100) facets, respectively.

Additional complete analyses of other representative example systems are presented in the SI Sec.~\ref{subsec:msmfitting} to illustrate how differences in geometry and surface coverage affect the equilibrium distribution of states, the TICA spaces, and the total flux. This includes large nanoparticles with low and high coverages (Figs.~\ref{fig:fluxDilute} and~\ref{fig:fluxSaturated}) and a small nanoparticle and slabs with moderate coverages (Figs.~\ref{fig:flux100}-\ref{fig:fluxSmallNP}). The relative ordering of the slow dynamical processes differs for different systems. For example, on a nanoparticle with dilute coverage, the slowest process is \ce{H2} association/dissociation (Fig.~\ref{fig:fluxDilute}), while at higher coverages, switching between (111)/(100) facets becomes the slowest process. Naturally, in the slab simulations there is only one type of facet and, consequently, no mode to swap between (100)/(111) facets. Interestingly, while the $\text{H}_{\text{top}}$ state emerged as a dynamically distinct state, it did not have any significant effect on the slowest dynamical modes, as can be seen in the flux analysis (Fig.~\ref{fig:flux} and Figs.~\ref{fig:fluxDilute}-\ref{fig:fluxSmallNP}), likely because the populations of these $\text{H}_{\text{top}}$ states are very low.

\subsubsection{Rates}
\begin{figure}
    \centering
    \includegraphics[width=0.75\linewidth]{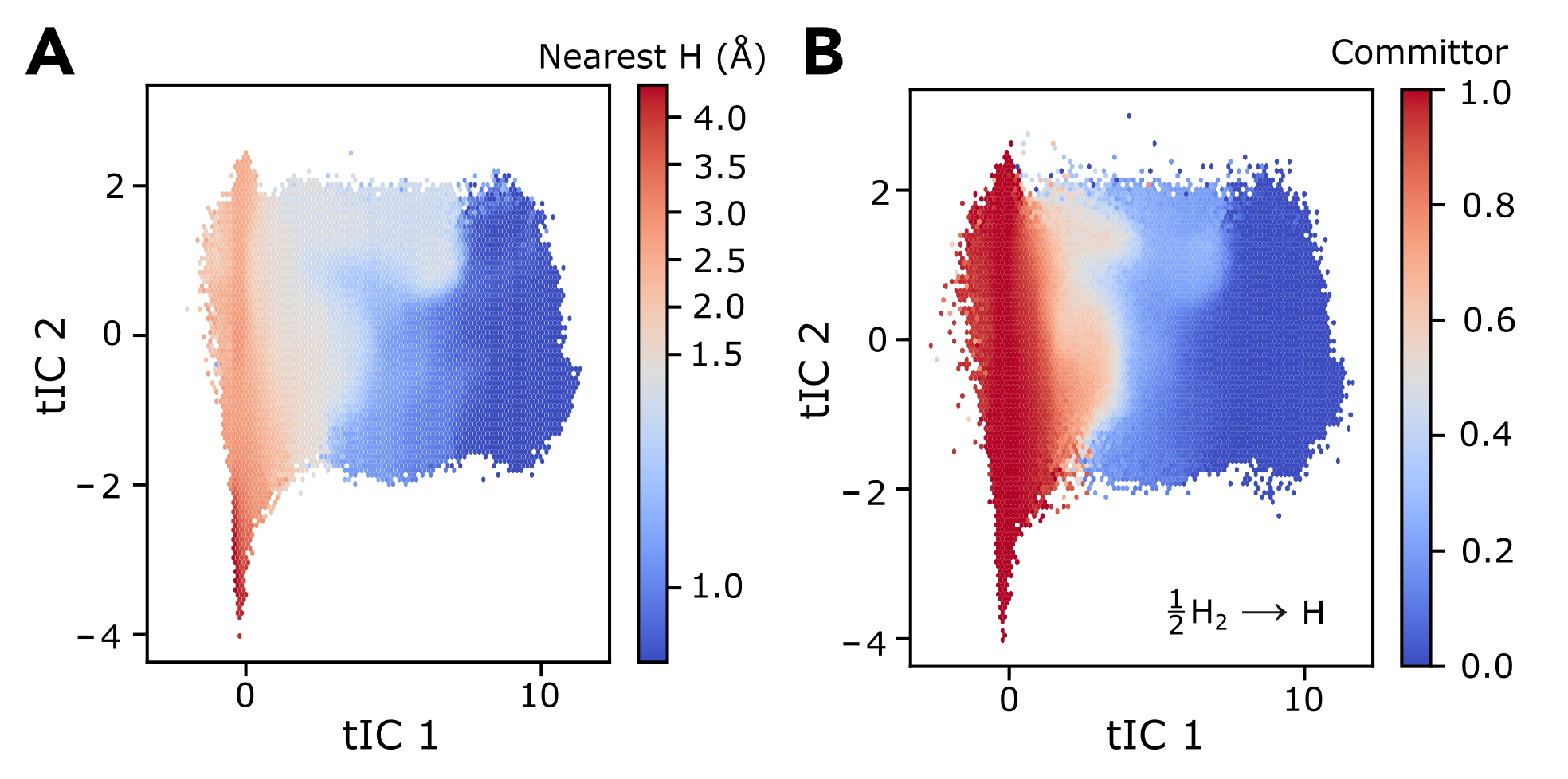}
    \caption{
    Configurations from  the simulation data described in Fig.~\ref{fig:flux} projected into TICA space and colored by A) the nearest distance to a neighboring hydrogen atom and B) the committor value for the dissociation reaction.}
    \label{fig:committor}
\end{figure}

In the Rh-H systems, the catalytic reaction of interest is the hydrogen dissociation and association reaction, and the rate of this reaction is a more measurable metric of catalytic activity than the timescale of the slowest contribution to this process. The rates of dissociation and association are calculated by defining configurations as reactant ($\bx \in A$), product ($\bx \in B$), and intermediate states ($\bx \notin A \cup B$) for these reactions and measuring the flux between them. The H and \ce{H2} reactant and product states are defined as having a nearest neighbor bond length greater than 1.5~\AA\ and less than 1.0~\AA, respectively, where the bond lengths are averaged over a moving frame of 50~fs (Fig.~\ref{fig:committor}A). To ensure that H$_{\text{top}}$ sites are not affecting the rate calculations, any hydrogen that has multiple nearest neighbors within a 1.5~\AA\ radius, as well as all of its neighbors, are classified as intermediate states.  
In a first step, the committor is estimated from the trajectory data.
The forward committor, $q^+(\bx)$, or probability of hitting the product state, $B$, before the reactant state, $A$, is given by
\begin{equation}
\label{eq:committor}
q^{+}(\bx) = \begin{cases}
      0 &  \bx \in A \\
      1 &  \bx \in B\\
      \mathcal{S}(\tau)q^+(\bx) &  \bx \notin A \cup B
    \end{cases} ,
\end{equation}
where $\mathcal{S}(\tau)$ is the transition operator with a stopping criterion in which the dynamics are terminated upon entering states $A$ or $B$.~\cite{strahan_long-time-scale_2021} 
The stopping criterion ensures that once a configuration enters the product state, it is committed to that state.
The committor is approximated from Eq.~\eqref{eq:committor} employing the dynamical Galerkin approximation (DGA)~\cite{thiede_galerkin_2019,strahan_long-time-scale_2021} with the indicator functions of intermediate microstates as basis functions for the committor expansion and solving the resulting set of linear equations (see SI Sec.~\ref{subsec:tpt} for details). 
The intermediate microstates are determined by  projecting all configurations $\bx \notin A \cup B$ into the TICA space and clustering into 64 discrete states using Kmeans++.
The resulting committor estimate $q^+(\bx)$ of configurations projected into TICA space is shown in Fig.~\ref{fig:committor}B.
If the system fulfills detailed balance, the forward committor, $q^+(\bx)$, is related to the backward committor, $q^-(\bx)$, 
by $q^+(\bx) = 1-q^-(\bx)$.~\cite{noe_constructing_2009} 
The rates are then calculated from the trajectory data for a lag time $\tau$ using the committor estimates to obtain the net reactive flux
%
%
\begin{equation}
\label{eq:rate}
F_{AB} = \frac{1}{\tau} \sum_{t=0}^{\tau-1} \mathbb{E}\left[q^-(\bx_{\max(0,S_t^-)})q^+(\bx_{\min(\tau,S_{t+1}^+)})
\big(q^+(\bx_{t+1}) - q^+(\bx_t)\big)
\right]
\end{equation}
and rates
\begin{equation}
 k_{AB} = \frac{F_{AB}}{\mathbb{E}[q^-(\bx)]}   \quad ,
\end{equation}
%
where the times $S_{t+1}^+$ and $S_t^-$ stop the system as soon as it enters states $A$ or $B$
(see SI Sec.~\ref{subsec:tpt} for details). \cite{strahan_long-time-scale_2021,lorpaiboon_exact_2026}
With this method, the calculation of the rate with the true committor should be independent of the choice of lag time.~\cite{lorpaiboon_exact_2026} Absolute values of the calculated rates had only a minor dependence on the definition of $A$ and $B$, indicating that appropriate bounds were chosen to distinguish the states. 

\subsection{Reactivity as a Function of Surface Coverage and Geometry}

The MSM analysis described above is applied to investigate the dynamical processes in the Rh-H systems over a range of different surface geometries and adsorbate coverages. Four geometries are included (Fig.~\ref{fig:MSMconstruction}A): slabs with (100) facet terminations, slabs with (111) facet terminations, and two sizes of cuboctahedral nanoparticles with (100) and (111) facet terminations (2~nm and 5~nm diameters). For each of these geometries, a range of hydrogen coverages is considered, from the dilute limit with about 0.05 monolayer (ML) coverage to the saturated limit with about 1.0 ML coverage of hydrogen (corresponding to $\leq$ 1 atm partial pressure of hydrogen). Trajectory data are generated by running MD simulations in the NVT ensemble at $T=450$~K 
accumulating 200~ns of hydrogen atom trajectories for each system setup.
Simulation details are included in the SI Sec.\ref{subsec:simsetting}.

\subsubsection{MSM Timescales of the Hydrogen Association and Dissociation Reaction}
\begin{figure}
    \centering
    \includegraphics[width=\linewidth]{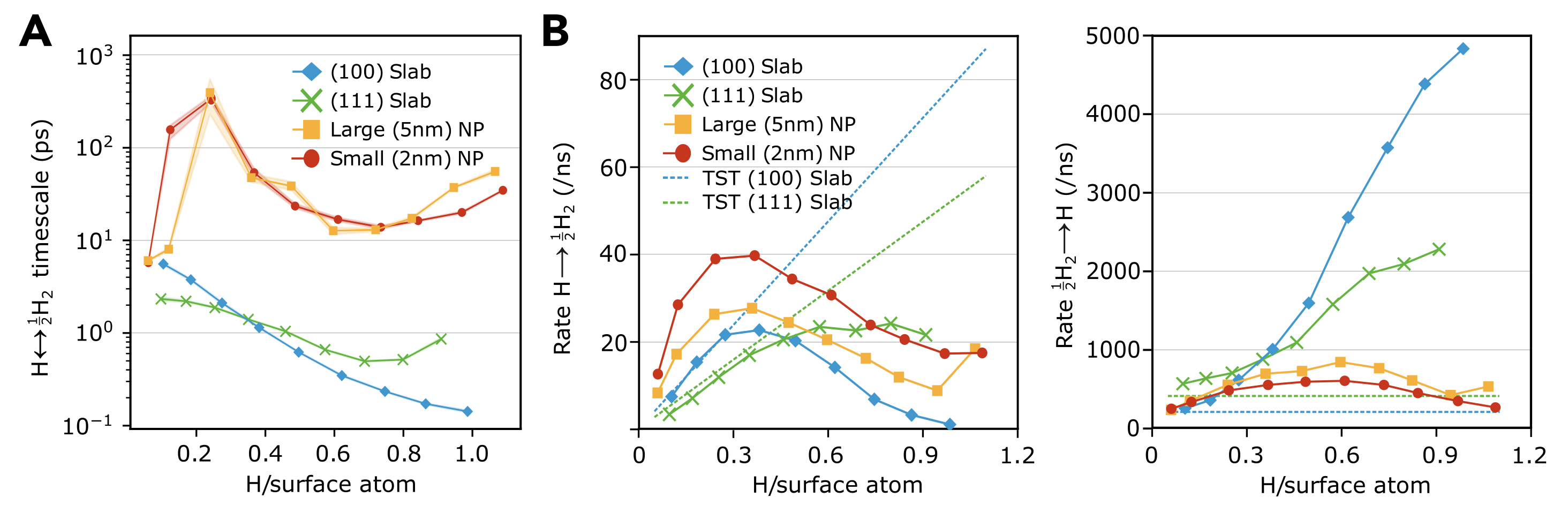}
    \caption{A) Timescales of the association/dissociation reactions on rhodium slabs and nanoparticles as a function of hydrogen concentration. B) Rates of the association (\ce{H}$\rightarrow\frac{1}{2}$\ce{H2}, left graph) and dissociation ($\frac{1}{2}$\ce{H2}$\rightarrow$\ce{H}, right graph) of hydrogen as a function of concentration. Dashed lines show TST rates on (100) and (111) surfaces.}
    \label{fig:rates}
\end{figure}

Focusing on the association/dissociation reaction of hydrogen on rhodium, we explore the relaxation timescales for these processes obtained from the eigenvalues of the MSMs as a function of geometry and hydrogen concentration, presented in Fig.~\ref{fig:rates}A.
While intuition would naively suggest that 
association/dissociation is slower on slabs than on nanoparticles, due to the presence of energetically unstable under-coordinated sites in the latter, our results reveal the opposite behavior. The slabs exhibit faster association/dissociation timescales over the whole range of hydrogen surface coverages.
At very low hydrogen concentrations, the timescales on the nanoparticles and slabs are comparable as the $\frac{1}{2}$\ce{H2}$\leftrightarrow$H flux is dominated by processes on the facets of the nanoparticles (see e.g. Fig.~\ref{fig:fluxDilute} in the SI). It matches the timescales on the (100) slab more closely since, in the dilute limit, the equilibrium concentration of hydrogen is much higher on the (100) facets of the nanoparticles as compared to the (111) facets.
For nanoparticles with hydrogen coverages greater than 10\%, the major contribution to the $\frac{1}{2}$\ce{H2}$\leftrightarrow$H flux is from \ce{H2} on corners and edges (see e.g. Figs.~\ref{fig:flux} and~\ref{fig:fluxSaturated} in the SI).
This indicates that the corner and edge sites function as traps for physisorbed \ce{H2} molecules, slowing down the overall reaction. 
The trend in the timescales as a function of hydrogen concentration is comparable for the small and large nanoparticles. The steep increase in timescales caused by the trapping on corners and edges is followed by a decrease due to the increasing availability of hydrogen atoms.
The (111) facet exhibits smaller timescales than the (100) facet in the dilute limit but at moderate surface coverage, dissociation/association becomes faster on the (100) facet. This trend is also observed in the relative rates of \ce{H2} association and disocciation which is discussed in more detail in the following section. 


To confirm that sufficient data was sampled for building the MSMs as well as to verify the assumption that any correlations between individual hydrogen trajectories from the same simulation do not affect the results, five MSMs were fitted for each system using 50\% of the hydrogen trajectories randomly selected.  
The error bars in Fig.~\ref{fig:rates}A represent the 95\% confidence interval over the timescales calculated from the five MSM fits,
validating that 100~ns (half of the total simulation data acquired) was sufficient to analyze the trends in the timescales. 
If the MSM is built from a single, long MD trajectory, 
the length of the required simulation time is proportional to the timescale of the reaction to be observed. 
For very slow reactions with high energy barriers that cannot be sampled directly, trajectory data may be collected from a series of simulations launched in various parts of the phase space. In some cases, biased simulations may be required that need to be reweighted to preserve the correct dynamics.~\cite{kieninger_dynamical_2020,keller_dynamical_2024,donati_girsanov_2018} 
For some of the studied systems, the timescale of association/dissociation was very close to that of another mode, which could lead to modes mixing, meaning that the two modes were not well separated. In those cases, the timescales were nearly degenerate and therefore had no effect on the resulting trends in timescales.

\subsubsection{Rates of Hydrogen Association and Dissociation}
In addition to the relaxation timescales, we explore the dependence of the rates of hydrogen association and dissociation on the rhodium surface geometries and hydrogen coverages, shown in Fig.~\ref{fig:rates}B. The rates were estimated from the trajectory data using Eq.~\eqref{eq:rate}. For a direct comparison with typical rate estimates on static surfaces, we compute the rates on the (100) and (111) surface in the dilute limit with standard TST, using the nudged elastic band (NEB) method~\cite{jonsson_nudged_1998,henkelman_climbing_2000} to determine the corresponding energy barriers (details regarding the NEB calculations can be found in SI Sec.~\ref{subsec:neb}). 
Assuming no interactions between adsorbates on the surface, the TST rate for H$\to\frac{1}{2}$\ce{H2} depends linearly on the surface coverage while the $\frac{1}{2}$\ce{H2}$\to$H rate is constant, shown as dashed lines in Fig.~\ref{fig:rates}B. 
In the dilute limit, the rate estimates from the trajectory data are consistent with the TST rates.
At higher hydrogen concentrations and for the different geometries, qualitatively different trends are observed due to the complex dynamical interplay of various processes that is not captured by standard TST. 
Comparing the rate estimates for \ce{H2} dissociation and association on slabs and nanoparticles, the geometric features (corners, edges) are found to have a pronounced effect on the rates.

The rates of hydrogen association (left graph in Fig.~\ref{fig:rates}B) peak before full coverage, which suggests  competing effects cause slow-downs at high and low coverages. At low coverages, the association rates on the (100) and (111) slabs increase linearly with concentration as pairs of hydrogen atoms are more readily available, which is also consistent with the TST predictions. At high coverages, TST would suggest a continuous increase of the association rates, whereas the rates extracted from the trajectory data start to decrease. Nearly all the hydrogen binding sites are filled, yet there are fewer \emph{productive} collisions. One hypothesis as to why the association rates slow at higher coverages is 
that hydrogen atoms on a saturated surface often encounter other hydrogens, but rarely possess sufficient momentum to overcome the energy barrier for association.
The \ce{H}$\rightarrow\frac{1}{2}$\ce{H2} reaction on the (100) surface appears to be more sensitive to this effect than it is on the (111) surface, leading to a crossover of association rates on the (111) and (100) surfaces for higher concentrations. 
A possible explanation for the different trends in the association rates on these two surfaces could be 
that the (111) surface has two hollow binding sites for every monolayer of adsorbed hydrogen. Consequently, at the same coverage, twice as many open hollow sites are available, making the (111) surface less crowded and diffusion more accessible.

Rates of hydrogen association on nanoparticles exhibit similar trends as on the (100) surface but are overall enhanced.
Binding to (100) facets is energetically more favorable than to (111) facets, meaning that hydrogen atoms on nanoparticles tend to concentrate on (100) facets, and the majority of the \ce{H}$\rightarrow\frac{1}{2}$\ce{H2} flux involves \ce{H_{2,(100)}} (see e.g. Fig.~\ref{fig:fluxDilute} in the SI). The enhanced rates of association may be a result of this accumulation of the hydrogen atoms to the (100) facets, leading to an effectively higher concentration, in combination with hydrogen atoms on edge sites. The difference in the rates is largest for the smaller nanoparticle, which is consistent with the fact that as the nanoparticle grows, the relative number of atoms on facets increases as compared to edges. 
The  nanoparticles with 2~nm and 5~nm diameters have 12.5/6.67/10 and 40.5/24/18 (100)/(111)/edge atoms per corner atom, respectively. 

The rates of hydrogen dissociation (right graph in Fig.~\ref{fig:rates}) are again in agreement with the TST rates in the dilute limit but show a steep increase with hydrogen concentration for the (100) and (111) surfaces.
Two effects are likely to enhance \ce{H2} dissociation as a function of hydrogen coverage: a lowering of the energy barrier for dissociation at higher concentrations and momentum transfer from diffusing H atoms.
The difference in \ce{H2} dissociation rates between the (100) and (111) surface at higher concentrations may be understood in terms of available adsorption sites during the dissociation process.
The lowest energy dissociation pathway on a (100) surface first places the two hydrogen atoms into neighboring bridge sites, which are higher in energy than the hollow sites and, therefore, typically unoccupied. In contrast, \ce{H2} dissociation on the (111) surface is not mediated by bridge sites and, consequently, empty adsorption sites are less readily available in the high concentration limit. Dissociation on nanoparticles is notably slower than on surfaces, especially at high concentrations of hydrogen. Likely, the dissociation of \ce{H2} from corner and edge sites is less affected by neighboring adsorbates and the high energy barrier to dissociate from these sites keeps the molecules trapped. For low concentrations, dissociation rates on the larger nanoparticle resembles the rates on the surfaces to a slightly greater extent than on the smaller nanoparticle due to the larger ratio of facet to edge/corner sites.  

Similar to the discussion of the relaxation timescale, the edge and corner sites of the nanoparticles do not provide preferred dissociation sites for \ce{H2} molecules. By contrast, these sites appear to be acting as dynamical traps leading to an overall slowdown of \ce{H2} dissociation on nanoparticles as compared to (100) and (111) surfaces.

\section{Conclusions}
We introduce Markov state models for catalytic reactions to disentangle and analyze the complex dynamical processes observed in these systems.
A typical heterogeneous catalyst operates at high temperatures and pressures and exhibits nanoscale features that can dynamically evolve over the course of the reaction. To capture the interdependence of the various processes, it is imperative that these realistic conditions are represented in the theoretical model. Applying MSMs to unbiased simulation trajectories of catalytic reactions provides insight into which factors will speed up, slow down, or poison the catalytic process. Notably, the approach does not necessitate defining the reactive events of interest, but rather identifies the slowest reaction modes automatically. In addition to giving reaction rates, MSMs provide relaxation timescales that characterize how the system approaches equilibrium, offering insight into dynamical bottlenecks that may not be apparent from rates alone.

We demonstrate the applicability of our MSM-based approach by studying a simple catalytic system involving hydrogen interacting with Rh surfaces, specifically with the goal of understanding the impact of surface features and \ce{H2} pressures on catalytic activity. 
The rates of \ce{H2} dissociation and association are indeed impacted by the geometric features present in nanoparticles, 
with \ce{H2} molecules becoming trapped at undercoordinated sites, effectively reducing the dissociation rate on nanoparticles as compared to slabs. At the same time, the association rate 
is accelerated by nanoscale features. The relaxation timescales extracted from the MSMs show that, overall, the equilibrium flux between H $\leftrightarrow\frac{1}{2}$\ce{H2} is less on nanoparticles than slabs. 
Contrary to intuitive expectations that  higher concentrations of reactants on the surface would lead to monotonic increases in the reaction rates, our analysis predicts a downturn in the \ce{H2} association rate with increasing surface saturation of hydrogen on nanoparticles and the (100) slab. 
The reduced association rate at high concentrations is most likely linked to the lower mobility of individual hydrogen atoms on an almost fully occupied surface.
This crowding phenomenon may have a greater effect on reactions with multiple reactant species, since the surface dynamics will be more complex than for hydrogen. 

Even for this seemingly simple catalytic system, the observed trends are partially unexpected, emphasizing the need to explore the full dynamical picture of complex catalytic systems under operando conditions to be able to uncover the unique behavior seen in catalysis at the nanoscale.
The analysis of the Rh-H system allows us to assess the usefulness of applying MSMs for other catalytic systems, as well as for other dynamical processes in condensed matter. Future challenges include the investigation of catalysts with multiple reacting elements which necessitates an explicit consideration of all relevant elements in the MSM. In addition,  the study of slow reactions with high energy barriers will likely require enhanced sampling techniques together with a careful unbiasing of the simulation data to retain the correct dynamics.

\section{Data Availability}
The Au-Rh-H potential is available in the SI and can be run in LAMMPS~\cite{thompson_lammps_2022} or with the Atomic Simulation Environment package\cite{thompson_lammps_2022,hjorth_larsen_atomic_2017} (pacemaker/pyace version: 0.2.7+163.g0ad96ce). Simulation trajectories are available upon request. 

The code to obtain the committor estimates and rate calculations is available at \url{https://github.com/flatiron-icc/tpt-utils}.

\section{Acknowledgements}
The Flatiron Institute is a division of the Simons Foundation.

\newpage
\bibliographystyle{achemso_jabbrv}
\bibliography{msm}

@article{keller_dynamical_2024,
	title = {Dynamical {Reweighting} for {Biased} {Rare} {Event} {Simulations}},
	volume = {75},
	copyright = {http://creativecommons.org/licenses/by/4.0/},
	issn = {0066-426X, 1545-1593},
	url = {https://www.annualreviews.org/content/journals/10.1146/annurev-physchem-083122-124538},
	doi = {10.1146/annurev-physchem-083122-124538},
	OPTlanguage = {en},
	number = {1},
	urldate = {2024-11-08},
	journal = {Annual Review of Physical Chemistry},
	author = {Keller, Bettina G. and Bolhuis, Peter G.},
	month = jun,
	year = {2024},
	note = {Number: 1},
	pages = {137--162},
}

@article{donati_girsanov_2018,
	title = {Girsanov reweighting for metadynamics simulations},
	volume = {149},
	issn = {0021-9606, 1089-7690},
	url = {https://pubs.aip.org/jcp/article/149/7/072335/1075425/Girsanov-reweighting-for-metadynamics-simulations},
	doi = {10.1063/1.5027728},
	OPTlanguage = {en},
	number = {7},
	urldate = {2024-12-16},
	journal = {The Journal of Chemical Physics},
	author = {Donati, Luca and Keller, Bettina G.},
	month = aug,
	year = {2018},
	note = {Number: 7},
	pages = {072335},
}

@article{kieninger_dynamical_2020,
	title = {Dynamical reweighting methods for {Markov} models},
	volume = {61},
	issn = {0959440X},
	url = {https://linkinghub.elsevier.com/retrieve/pii/S0959440X19301575},
	doi = {10.1016/j.sbi.2019.12.018},
	OPTlanguage = {en},
	urldate = {2024-11-08},
	journal = {Current Opinion in Structural Biology},
	author = {Kieninger, Stefanie and Donati, Luca and Keller, Bettina G},
	month = apr,
	year = {2020},
	pages = {124--131},
}

@article{arbon_markov_2024,
	title = {Markov {State} {Models}: {To} {Optimize} or {Not} to {Optimize}},
	volume = {20},
	copyright = {https://creativecommons.org/licenses/by/4.0/},
	issn = {1549-9618, 1549-9626},
	shorttitle = {Markov {State} {Models}},
	url = {https://pubs.acs.org/doi/10.1021/acs.jctc.3c01134},
	doi = {10.1021/acs.jctc.3c01134},
	OPTlanguage = {en},
	number = {2},
	urldate = {2024-11-11},
	journal = {Journal of Chemical Theory and Computation},
	author = {Arbon, Robert E. and Zhu, Yanchen and Mey, Antonia S. J. S.},
	month = jan,
	year = {2024},
	note = {Number: 2},
	pages = {977--988},
}

@article{xie_graph_2019,
	title = {Graph dynamical networks for unsupervised learning of atomic scale dynamics in materials},
	volume = {10},
	issn = {2041-1723},
	url = {https://www.nature.com/articles/s41467-019-10663-6},
	doi = {10.1038/s41467-019-10663-6},
	OPTlanguage = {en},
	number = {1},
	urldate = {2024-11-11},
	journal = {Nature Communications},
	author = {Xie, Tian and France-Lanord, Arthur and Wang, Yanming and Shao-Horn, Yang and Grossman, Jeffrey C.},
	month = jun,
	year = {2019},
	note = {Number: 1},
	pages = {2667},
}

@article{husic_markov_2018,
	title = {Markov {State} {Models}: {From} an {Art} to a {Science}},
	volume = {140},
	issn = {0002-7863, 1520-5126},
	shorttitle = {Markov {State} {Models}},
	url = {https://pubs.acs.org/doi/10.1021/jacs.7b12191},
	doi = {10.1021/jacs.7b12191},
	OPTlanguage = {en},
	number = {7},
	urldate = {2024-12-08},
	journal = {Journal of the American Chemical Society},
	author = {Husic, Brooke E. and Pande, Vijay S.},
	month = feb,
	year = {2018},
	note = {Number: 7},
	pages = {2386--2396},
}

@article{perez-hernandez_identification_2013,
	title = {Identification of slow molecular order parameters for {Markov} model construction},
	volume = {139},
	issn = {0021-9606, 1089-7690},
	url = {https://pubs.aip.org/jcp/article/139/1/015102/192538/Identification-of-slow-molecular-order-parameters},
	doi = {10.1063/1.4811489},
	OPTlanguage = {en},
	number = {1},
	urldate = {2025-01-06},
	journal = {The Journal of Chemical Physics},
	author = {Pérez-Hernández, Guillermo and Paul, Fabian and Giorgino, Toni and De Fabritiis, Gianni and Noé, Frank},
	month = jul,
	year = {2013},
	pages = {015102},
}

@article{prinz_markov_2011,
	title = {Markov models of molecular kinetics: {Generation} and validation},
	volume = {134},
	issn = {0021-9606, 1089-7690},
	shorttitle = {Markov models of molecular kinetics},
	url = {https://pubs.aip.org/jcp/article/134/17/174105/699460/Markov-models-of-molecular-kinetics-Generation-and},
	doi = {10.1063/1.3565032},
	OPTlanguage = {en},
	number = {17},
	urldate = {2025-01-06},
	journal = {The Journal of Chemical Physics},
	author = {Prinz, Jan-Hendrik and Wu, Hao and Sarich, Marco and Keller, Bettina and Senne, Martin and Held, Martin and Chodera, John D. and Schütte, Christof and Noé, Frank},
	month = may,
	year = {2011},
	pages = {174105},
}

@article{noe_constructing_2009,
	title = {Constructing the equilibrium ensemble of folding pathways from short off-equilibrium simulations},
	volume = {106},
	issn = {0027-8424, 1091-6490},
	url = {https://pnas.org/doi/full/10.1073/pnas.0905466106},
	doi = {10.1073/pnas.0905466106},
	OPTlanguage = {en},
	number = {45},
	urldate = {2025-06-24},
	journal = {Proceedings of the National Academy of Sciences},
	author = {Noé, Frank and Schütte, Christof and Vanden-Eijnden, Eric and Reich, Lothar and Weikl, Thomas R.},
	month = nov,
	year = {2009},
	pages = {19011--19016},
}

@article{prinz_probing_2011,
	title = {Probing molecular kinetics with {Markov} models: metastable states, transition pathways and spectroscopic observables},
	volume = {13},
	issn = {1463-9076, 1463-9084},
	shorttitle = {Probing molecular kinetics with {Markov} models},
	url = {https://xlink.rsc.org/?DOI=c1cp21258c},
	doi = {10.1039/c1cp21258c},
	OPTlanguage = {en},
	number = {38},
	urldate = {2025-06-24},
	journal = {Physical Chemistry Chemical Physics},
	author = {Prinz, Jan-Hendrik and Keller, Bettina and Noé, Frank},
	year = {2011},
	pages = {16912},
}

@article{li_unveiling_2025,
	title = {Unveiling hidden reaction kinetics of carbon dioxide in supercritical aqueous solutions},
	volume = {122},
	url = {https://www.pnas.org/doi/10.1073/pnas.2406356121},
	doi = {10.1073/pnas.2406356121},
	number = {1},
	urldate = {2025-09-03},
	journal = {Proceedings of the National Academy of Sciences},
	author = {Li, Chu and Yao, Yuan and Pan, Ding},
	month = jan,
	year = {2025},
	note = {Publisher: Proceedings of the National Academy of Sciences},
	pages = {e2406356121},
}

@article{harrigan_msmbuilder_2017,
	title = {{MSMBuilder}: {Statistical} {Models} for {Biomolecular} {Dynamics}},
	volume = {112},
	issn = {0006-3495},
	shorttitle = {{MSMBuilder}},
	url = {https://www.sciencedirect.com/science/article/pii/S0006349516310487},
	doi = {10.1016/j.bpj.2016.10.042},
	number = {1},
	urldate = {2025-09-04},
	journal = {Biophysical Journal},
	author = {Harrigan, Matthew P. and Sultan, Mohammad M. and Hernández, Carlos X. and Husic, Brooke E. and Eastman, Peter and Schwantes, Christian R. and Beauchamp, Kyle A. and McGibbon, Robert T. and Pande, Vijay S.},
	month = jan,
	year = {2017},
	pages = {10--15},
}

@article{wang_constructing_2018,
	title = {Constructing {Markov} {State} {Models} to elucidate the functional conformational changes of complex biomolecules},
	volume = {8},
	copyright = {© 2017 Wiley Periodicals, Inc.},
	issn = {1759-0884},
	url = {https://onlinelibrary.wiley.com/doi/abs/10.1002/wcms.1343},
	doi = {10.1002/wcms.1343},
	OPTlanguage = {en},
	number = {1},
	urldate = {2025-11-20},
	journal = {WIREs Computational Molecular Science},
	author = {Wang, Wei and Cao, Siqin and Zhu, Lizhe and Huang, Xuhui},
	year = {2018},
	note = {\_eprint: https://wires.onlinelibrary.wiley.com/doi/pdf/10.1002/wcms.1343},
	pages = {e1343},
}

@article{chen_complete_2024,
	title = {Complete miscibility of immiscible elements at the nanometre scale},
	volume = {19},
	copyright = {2024 The Author(s), under exclusive licence to Springer Nature Limited},
	issn = {1748-3395},
	url = {https://www.nature.com/articles/s41565-024-01626-0},
	doi = {10.1038/s41565-024-01626-0},
	Optlanguage = {en},
	number = {6},
	urldate = {2024-06-20},
	journal = {Nature Nanotechnology},
	author = {Chen, Peng-Cheng and Gao, Mengyu and McCandler, Caitlin A. and Song, Chengyu and Jin, Jianbo and Yang, Yao and Maulana, Arifin Luthfi and Persson, Kristin A. and Yang, Peidong},
	month = jun,
	year = {2024},
	note = {Publisher: Nature Publishing Group},
	pages = {775--781},
}

@article{perdew_generalized_1996,
	title = {Generalized {Gradient} {Approximation} {Made} {Simple}},
	volume = {77},
	url = {https://link.aps.org/doi/10.1103/PhysRevLett.77.3865},
	doi = {10.1103/PhysRevLett.77.3865},
	number = {18},
	urldate = {2022-03-10},
	journal = {Physical Review Letters},
	author = {Perdew, John P. and Burke, Kieron and Ernzerhof, Matthias},
	month = oct,
	year = {1996},
	note = {Publisher: American Physical Society},
	pages = {3865--3868},
}

@article{kresse_ultrasoft_1999,
	title = {From ultrasoft pseudopotentials to the projector augmented-wave method},
	volume = {59},
	url = {https://link.aps.org/doi/10.1103/PhysRevB.59.1758},
	doi = {10.1103/PhysRevB.59.1758},
	number = {3},
	urldate = {2022-03-10},
	journal = {Physical Review B},
	author = {Kresse, G. and Joubert, D.},
	month = jan,
	year = {1999},
	note = {Publisher: American Physical Society},
	pages = {1758--1775},
}

@article{kresse_efficient_1996,
	title = {Efficient iterative schemes for ab initio total-energy calculations using a plane-wave basis set},
	volume = {54},
	url = {https://link.aps.org/doi/10.1103/PhysRevB.54.11169},
	doi = {10.1103/PhysRevB.54.11169},
	number = {16},
	urldate = {2022-03-10},
	journal = {Physical Review B},
	author = {Kresse, G. and Furthmüller, J.},
	month = oct,
	year = {1996},
	note = {Publisher: American Physical Society},
	pages = {11169--11186},
}

@article{lysogorskiy_performant_2021,
	title = {Performant implementation of the atomic cluster expansion ({PACE}) and application to copper and silicon},
	volume = {7},
	copyright = {2021 The Author(s)},
	issn = {2057-3960},
	url = {https://www.nature.com/articles/s41524-021-00559-9},
	doi = {10.1038/s41524-021-00559-9},
	Optlanguage = {en},
	number = {1},
	urldate = {2024-05-23},
	journal = {npj Computational Materials},
	author = {Lysogorskiy, Yury and Oord, Cas van der and Bochkarev, Anton and Menon, Sarath and Rinaldi, Matteo and Hammerschmidt, Thomas and Mrovec, Matous and Thompson, Aidan and Csányi, Gábor and Ortner, Christoph and Drautz, Ralf},
	month = jun,
	year = {2021},
	note = {Publisher: Nature Publishing Group},
	pages = {1--12},
}

@article{lysogorskiy_active_2023,
	title = {Active learning strategies for atomic cluster expansion models},
	volume = {7},
	url = {https://link.aps.org/doi/10.1103/PhysRevMaterials.7.043801},
	doi = {10.1103/PhysRevMaterials.7.043801},
	number = {4},
	urldate = {2024-03-03},
	journal = {Physical Review Materials},
	author = {Lysogorskiy, Yury and Bochkarev, Anton and Mrovec, Matous and Drautz, Ralf},
	month = apr,
	year = {2023},
	note = {Publisher: American Physical Society},
	pages = {043801},
}

@article{drautz_atomic_2019,
	title = {Atomic cluster expansion for accurate and transferable interatomic potentials},
	volume = {99},
	issn = {2469-9950, 2469-9969},
	url = {https://link.aps.org/doi/10.1103/PhysRevB.99.014104},
	doi = {10.1103/PhysRevB.99.014104},
	Optlanguage = {en},
	number = {1},
	urldate = {2021-04-23},
	journal = {Physical Review B},
	author = {Drautz, Ralf},
	month = jan,
	year = {2019},
	pages = {014104},
}

@article{mccormack_mechanisms_2005,
	title = {Mechanisms of {H2} dissociative adsorption on the {Pt}(211) stepped surface},
	volume = {122},
	issn = {0021-9606, 1089-7690},
	url = {https://pubs.aip.org/jcp/article/122/19/194708/973365/Mechanisms-of-H2-dissociative-adsorption-on-the-Pt},
	doi = {10.1063/1.1900087},
	Optlanguage = {en},
	number = {19},
	urldate = {2024-10-16},
	journal = {The Journal of Chemical Physics},
	author = {McCormack, Drew A. and Olsen, Roar A. and Baerends, Evert Jan},
	month = may,
	year = {2005},
	pages = {194708},
}

@article{ledentu_ab_1998,
	title = {Ab initio study of the dissociative adsorption of {H2} on the {Pd}(110) surface},
	volume = {412-413},
	copyright = {https://www.elsevier.com/tdm/userlicense/1.0/},
	issn = {00396028},
	url = {https://linkinghub.elsevier.com/retrieve/pii/S0039602898004828},
	doi = {10.1016/S0039-6028(98)00482-8},
	Optlanguage = {en},
	urldate = {2024-10-16},
	journal = {Surface Science},
	author = {Ledentu, V. and Dong, W. and Sautet, P.},
	month = sep,
	year = {1998},
	pages = {518--526},
}

@article{feibelman_orientation_1991,
	title = {Orientation dependence of the hydrogen molecule’s interaction with {Rh}(001)},
	volume = {67},
	copyright = {http://link.aps.org/licenses/aps-default-license},
	issn = {0031-9007},
	url = {https://link.aps.org/doi/10.1103/PhysRevLett.67.461},
	doi = {10.1103/PhysRevLett.67.461},
	Optlanguage = {en},
	number = {4},
	urldate = {2024-10-16},
	journal = {Physical Review Letters},
	author = {Feibelman, Peter J.},
	month = jul,
	year = {1991},
	pages = {461--464},
}

@article{eichler_ab-initio_1998,
	title = {Ab-initio calculations of the {6D} potential energy surfaces for the dissociative adsorption of {H2} on the (100) surfaces of {Rh}, {Pd} and {Ag}},
	volume = {397},
	copyright = {https://www.elsevier.com/tdm/userlicense/1.0/},
	issn = {00396028},
	url = {https://linkinghub.elsevier.com/retrieve/pii/S0039602897007243},
	doi = {10.1016/S0039-6028(97)00724-3},
	Optlanguage = {en},
	number = {1-3},
	urldate = {2024-10-16},
	journal = {Surface Science},
	author = {Eichler, A. and Kresse, G. and Hafner, J.},
	month = feb,
	year = {1998},
	pages = {116--136},
}

@article{eichler_quantum_1996,
	title = {Quantum {Steering} {Effects} in the {Dissociative} {Adsorption} of {H2} on {Rh}(100)},
	volume = {77},
	url = {https://link.aps.org/doi/10.1103/PhysRevLett.77.1119},
	doi = {10.1103/PhysRevLett.77.1119},
	number = {6},
	urldate = {2024-10-16},
	journal = {Physical Review Letters},
	author = {Eichler, A. and Kresse, G. and Hafner, J.},
	month = aug,
	year = {1996},
	note = {Publisher: American Physical Society},
	pages = {1119--1122},
}

@article{stukowski_visualization_2009,
	title = {Visualization and analysis of atomistic simulation data with {OVITO}–the {Open} {Visualization} {Tool}},
	volume = {18},
	issn = {0965-0393},
	url = {https://doi.org/10.1088/0965-0393/18/1/015012},
	doi = {10.1088/0965-0393/18/1/015012},
	Optlanguage = {en},
	number = {1},
	urldate = {2025-10-09},
	journal = {Modelling and Simulation in Materials Science and Engineering},
	author = {Stukowski, Alexander},
	month = dec,
	year = {2009},
	pages = {015012},
}

@article{hjorth_larsen_atomic_2017,
	title = {The atomic simulation environment—a {Python} library for working with atoms},
	volume = {29},
	issn = {0953-8984},
	url = {https://doi.org/10.1088/1361-648X/aa680e},
	doi = {10.1088/1361-648X/aa680e},
	Optlanguage = {en},
	number = {27},
	urldate = {2025-10-09},
	journal = {Journal of Physics: Condensed Matter},
	author = {Hjorth Larsen, Ask and Jørgen Mortensen, Jens and Blomqvist, Jakob and Castelli, Ivano E and Christensen, Rune and Dułak, Marcin and Friis, Jesper and Groves, Michael N and Hammer, Bjørk and Hargus, Cory and Hermes, Eric D and Jennings, Paul C and Bjerre Jensen, Peter and Kermode, James and Kitchin, John R and Leonhard Kolsbjerg, Esben and Kubal, Joseph and Kaasbjerg, Kristen and Lysgaard, Steen and Bergmann Maronsson, Jón and Maxson, Tristan and Olsen, Thomas and Pastewka, Lars and Peterson, Andrew and Rostgaard, Carsten and Schiøtz, Jakob and Schütt, Ole and Strange, Mikkel and Thygesen, Kristian S and Vegge, Tejs and Vilhelmsen, Lasse and Walter, Michael and Zeng, Zhenhua and Jacobsen, Karsten W},
	month = jun,
	year = {2017},
	note = {Publisher: IOP Publishing},
	pages = {273002},
}

@article{thompson_lammps_2022,
	title = {{LAMMPS} - a flexible simulation tool for particle-based materials modeling at the atomic, meso, and continuum scales},
	volume = {271},
	issn = {0010-4655},
	url = {https://www.sciencedirect.com/science/article/pii/S0010465521002836},
	doi = {10.1016/j.cpc.2021.108171},
	urldate = {2025-10-09},
	journal = {Computer Physics Communications},
	author = {Thompson, Aidan P. and Aktulga, H. Metin and Berger, Richard and Bolintineanu, Dan S. and Brown, W. Michael and Crozier, Paul S. and in 't Veld, Pieter J. and Kohlmeyer, Axel and Moore, Stan G. and Nguyen, Trung Dac and Shan, Ray and Stevens, Mark J. and Tranchida, Julien and Trott, Christian and Plimpton, Steven J.},
	month = feb,
	year = {2022},
	pages = {108171},
}

@article{strahan_long-time-scale_2021,
	title = {Long-{Time}-{Scale} {Predictions} from {Short}-{Trajectory} {Data}: {A} {Benchmark} {Analysis} of the {Trp}-{Cage} {Miniprotein}},
	copyright = {© 2021 American Chemical Society},
	shorttitle = {Long-{Time}-{Scale} {Predictions} from {Short}-{Trajectory} {Data}},
	url = {https://pubs.acs.org/doi/full/10.1021/acs.jctc.0c00933},
	doi = {10.1021/acs.jctc.0c00933},
	Optlanguage = {en},
	urldate = {2025-10-28},
	journal = {Journal of Chemical Theory and Computation},
	author = {Strahan, John and Antoszewski, Adam and Lorpaiboon, Chatipat and Vani, Bodhi P. and Weare, Jonathan and Dinner, Aaron R.},
	month = apr,
	year = {2021},
	note = {Publisher: American Chemical Society},
}

@article{lai_automatic_2025,
	title = {Automatic {Process} {Exploration} through {Machine} {Learning} {Assisted} {Transition} {State} {Searches}},
	volume = {134},
	url = {https://link.aps.org/doi/10.1103/PhysRevLett.134.096201},
	doi = {10.1103/PhysRevLett.134.096201},
    OPTlanguage = {English},
	number = {9},
	urldate = {2026-02-10},
	journal = {Physical Review Letters},
	publisher = {American Physical Society},
	author = {Lai, King Chun and Poths, Patricia and Matera, Sebastian and Scheurer, Christoph and Reuter, Karsten},
	month = mar,
	year = {2025},
	pages = {096201},
}

@article{poths_ml-accelerated_2025,
	title = {{ML}-{Accelerated} {Automatic} {Process} {Exploration} {Reveals} {Facile} {O}-{Induced} {Pd} {Step}-{Edge} {Restructuring} on {Catalytic} {Time} {Scales}},
	volume = {15},
	url = {https://doi.org/10.1021/acscatal.4c06414},
	doi = {10.1021/acscatal.4c06414},
	OPTlanguage = {English},
	number = {1},
	urldate = {2025-11-19},
	journal = {ACS Catalysis},
	author = {Poths, Patricia and Lai, King Chun and Cannizzaro, Francesco and Scheurer, Christoph and Matera, Sebastian and Reuter, Karsten},
	month = jan,
	year = {2025},
	note = {Publisher: American Chemical Society},
	pages = {514--522},
}

@article{latz_three-dimensional_2012,
	title = {A three-dimensional self-learning kinetic {Monte} {Carlo} model: application to {Ag}(111)},
	volume = {24},
	issn = {0953-8984},
	shorttitle = {A three-dimensional self-learning kinetic {Monte} {Carlo} model},
	url = {https://doi.org/10.1088/0953-8984/24/48/485005},
	doi = {10.1088/0953-8984/24/48/485005},
	Optlanguage = {en},
	number = {48},
	urldate = {2025-11-19},
	journal = {Journal of Physics: Condensed Matter},
	author = {Latz, Andreas and Brendel, Lothar and Wolf, Dietrich E},
	month = oct,
	year = {2012},
	note = {Publisher: IOP Publishing},
	pages = {485005},
}

@article{vineyard_frequency_1957,
	title = {Frequency factors and isotope effects in solid state rate processes},
	volume = {3},
	issn = {0022-3697},
	url = {https://www.sciencedirect.com/science/article/pii/0022369757900598},
	doi = {10.1016/0022-3697(57)90059-8},
	Optlanguage = {en},
	number = {1},
	urldate = {2026-01-20},
	journal = {Journal of Physics and Chemistry of Solids},
	author = {Vineyard, George H.},
	month = jan,
	year = {1957},
	pages = {121--127},
}

@misc{zhang_non-equilibrium_2025,
	title = {Non-{Equilibrium} {Restructurings} in {Catalysis}: {A} {Chemical} {Space} {Odyssey}},
	shorttitle = {Non-{Equilibrium} {Restructurings} in {Catalysis}},
	url = {https://chemrxiv.org/engage/chemrxiv/article-details/69201e4265a54c2d4ae44325},
	doi = {10.26434/chemrxiv-2025-23xth-v2},
	urldate = {2026-01-14},
	publisher = {ChemRxiv},
	author = {Zhang, Zisheng and Zhou, Xuening},
	month = nov,
	year = {2025},
}

@article{reuter_ab_2016,
	address = {Dordrecht, Netherlands},
	title = {Ab {Initio} {Thermodynamics} and {First}-{Principles} {Microkinetics} for {Surface} {Catalysis}},
	volume = {146},
	copyright = {Catalysis Letters is a copyright of Springer, (2016). All Rights Reserved.},
	issn = {1011372X},
	url = {https://www.proquest.com/docview/2258884549/abstract/7AE77BD0015847B0PQ/1},
	doi = {10.1007/s10562-015-1684-3},
	Optlanguage = {en},
	number = {3},
	urldate = {2026-01-20},
	journal = {Catalysis Letters},
	publisher = {Springer Nature B.V.},
	author = {Reuter, Karsten},
	month = mar,
	year = {2016},
	note = {Num Pages: 541-563},
	pages = {541--563},
}

@article{sabbe_first-principles_2012,
	title = {First-principles kinetic modeling in heterogeneous catalysis: an industrial perspective on best-practice, gaps and needs},
	volume = {2},
	issn = {2044-4753, 2044-4761},
	shorttitle = {First-principles kinetic modeling in heterogeneous catalysis},
	url = {https://xlink.rsc.org/?DOI=c2cy20261a},
	doi = {10.1039/c2cy20261a},
	Optlanguage = {en},
	number = {10},
	urldate = {2026-01-20},
	journal = {Catalysis Science \& Technology},
	author = {Sabbe, Maarten K. and Reyniers, Marie-Françoise and Reuter, Karsten},
	year = {2012},
	pages = {2010},
}

@article{bonati_role_2023,
	title = {The role of dynamics in heterogeneous catalysis: {Surface} diffusivity and {N2} decomposition on {Fe}(111)},
	volume = {120},
	shorttitle = {The role of dynamics in heterogeneous catalysis},
	url = {https://www.pnas.org/doi/epub/10.1073/pnas.2313023120},
	doi = {10.1073/pnas.2313023120},
	Optlanguage = {en},
	number = {50},
	urldate = {2024-09-12},
	journal = {PNAS},
	author = {Bonati, Luigi and Polino, Daniela and Pizzolitto, Cristina and Biasi, Pierdomenico and Eckert, Rene and Reitmeier, Stephan and Schlögl, Robert and Michele, Parrinello},
	year = {2023},
	doi = {10.1073/pnas.2313023120},
}

@article{lorpaiboon_exact_2026,
	title = {An {Exact} {Multiple}-{Time}-{Step} {Variational} {Formulation} for the {Committor} and the {Transition} {Rate}},
	volume = {130},
	issn = {1520-6106},
	url = {https://doi.org/10.1021/acs.jpcb.5c06047},
	doi = {10.1021/acs.jpcb.5c06047},
	Optlanguage = {en},
	number = {1},
	urldate = {2026-01-20},
	journal = {The Journal of Physical Chemistry B},
	publisher = {American Chemical Society},
	author = {Lorpaiboon, Chatipat and Weare, Jonathan and Dinner, Aaron R.},
	month = jan,
	year = {2026},
	pages = {155--170},
}

@article{brunton_modern_2022,
	title = {Modern {Koopman} {Theory} for {Dynamical} {Systems}},
	volume = {64},
	issn = {0036-1445},
	url = {https://epubs.siam.org/doi/doi/10.1137/21M1401243},
	doi = {10.1137/21M1401243},
	number = {2},
	Optlanguage = {en},
	urldate = {2026-01-20},
	journal = {SIAM Review},
	publisher = {Society for Industrial and Applied Mathematics},
	author = {Brunton, Steven L. and Budišić, Marko and Kaiser, Eurika and Kutz, J. Nathan},
	month = may,
	year = {2022},
	pages = {229--340},
}

@inproceedings{arthur_k-means_2007,
	address = {USA},
	series = {{SODA} '07},
	title = {k-means++: the advantages of careful seeding},
	isbn = {978-0-89871-624-5},
	shorttitle = {k-means++},
	url = {https://dl.acm.org/doi/10.5555/1283383.1283494},
	urldate = {2026-01-20},
	booktitle = {Proceedings of the eighteenth annual {ACM}-{SIAM} symposium on {Discrete} algorithms},
	publisher = {Society for Industrial and Applied Mathematics},
	author = {Arthur, David and Vassilvitskii, Sergei},
	month = jan,
	year = {2007},
	pages = {1027--1035},
}

@article{trushin_self-learning_2005,
	title = {Self-learning kinetic {Monte} {Carlo} method: {Application} to {Cu}(111)},
	volume = {72},
	shorttitle = {Self-learning kinetic {Monte} {Carlo} method},
	url = {https://link.aps.org/doi/10.1103/PhysRevB.72.115401},
	doi = {10.1103/PhysRevB.72.115401},
	Optlanguage = {en},
	number = {11},
	urldate = {2026-01-20},
	journal = {Physical Review B},
	publisher = {American Physical Society},
	author = {Trushin, Oleg and Karim, Altaf and Kara, Abdelkader and Rahman, Talat S.},
	month = sep,
	year = {2005},
	pages = {115401},
}

@incollection{jonsson_nudged_1998,
	title = {Nudged elastic band method for finding minimum energy paths of transitions},
	isbn = {978-981-02-3498-0},
	url = {https://www.worldscientific.com/doi/abs/10.1142/9789812839664_0016},
	doi = {10.1142/9789812839664_0016},
	urldate = {2026-01-21},
	booktitle = {Classical and {Quantum} {Dynamics} in {Condensed} {Phase} {Simulations}},
	publisher = {World Scientific},
	author = {Jónsson, Hannes and Mills, Greg and Jacobsen, Karsten W.},
	month = jun,
	year = {1998},
	pages = {385--404},
}

@article{henkelman_climbing_2000,
	title = {A climbing image nudged elastic band method for finding saddle points and minimum energy paths},
	volume = {113},
	issn = {0021-9606},
	url = {https://doi.org/10.1063/1.1329672},
	doi = {10.1063/1.1329672},
	Optlanguage = {en},
	number = {22},
	urldate = {2026-01-21},
	journal = {The Journal of Chemical Physics},
	author = {Henkelman, Graeme and Uberuaga, Blas P. and Jónsson, Hannes},
	month = dec,
	year = {2000},
	pages = {9901--9904},
}

@article{thiede_galerkin_2019,
	title = {Galerkin approximation of dynamical quantities using trajectory data},
	volume = {150},
	issn = {0021-9606},
	url = {https://doi.org/10.1063/1.5063730},
	doi = {10.1063/1.5063730},
	Optlanguage = {en},
	number = {24},
	urldate = {2026-01-22},
	journal = {The Journal of Chemical Physics},
	author = {Thiede, Erik H. and Giannakis, Dimitrios and Dinner, Aaron R. and Weare, Jonathan},
	month = jun,
	year = {2019},
	pages = {244111},
}

@article{tao_situ_2011,
	title = {In {Situ} {Studies} of {Chemistry} and {Structure} of {Materials} in {Reactive} {Environments}},
	volume = {331},
	url = {https://www.science.org/doi/10.1126/science.1197461},
	doi = {10.1126/science.1197461},
	Optlanguage = {en},
	number = {6014},
	urldate = {2025-01-03},
	journal = {Science},
	publisher = {American Association for the Advancement of Science},
	author = {Tao, Franklin (Feng) and Salmeron, Miquel},
	month = jan,
	year = {2011},
	pages = {171--174},
}

@article{el-mellouhi_kinetic_2008,
	title = {Kinetic activation-relaxation technique: {An} off-lattice self-learning kinetic {Monte} {Carlo} algorithm},
	volume = {78},
	shorttitle = {Kinetic activation-relaxation technique},
	url = {https://link.aps.org/doi/10.1103/PhysRevB.78.153202},
	doi = {10.1103/PhysRevB.78.153202},
	Optlanguage = {en},
	number = {15},
	urldate = {2026-01-22},
	journal = {Physical Review B},
	publisher = {American Physical Society},
	author = {El-Mellouhi, Fedwa and Mousseau, Normand and Lewis, Laurent J.},
	month = oct,
	year = {2008},
	pages = {153202},
}

@article{xu_adaptive_2008,
	title = {Adaptive kinetic {Monte} {Carlo} for first-principles accelerated dynamics},
	volume = {129},
	issn = {0021-9606},
	url = {https://doi.org/10.1063/1.2976010},
	doi = {10.1063/1.2976010},
	Optlanguage = {en},
	number = {11},
	urldate = {2026-01-22},
	journal = {The Journal of Chemical Physics},
	author = {Xu, Lijun and Henkelman, Graeme},
	month = sep,
	year = {2008},
	pages = {114104},
}

@article{molgedey_separation_1994,
	title = {Separation of a mixture of independent signals using time delayed correlations},
	volume = {72},
	copyright = {http://link.aps.org/licenses/aps-default-license},
	issn = {0031-9007},
	url = {https://link.aps.org/doi/10.1103/PhysRevLett.72.3634},
	doi = {10.1103/PhysRevLett.72.3634},
	Optlanguage = {en},
	number = {23},
	urldate = {2026-02-09},
	journal = {Physical Review Letters},
	author = {Molgedey, L. and Schuster, H. G.},
	month = jun,
	year = {1994},
	pages = {3634--3637},
}

\setcounter{table}{0}
\setcounter{figure}{0}
\renewcommand{\thefigure}{S\arabic{figure}}
\renewcommand{\theequation}{S\arabic{equation}}
\renewcommand{\thetable}{S\arabic{table}}
\setcounter{equation}{0}
\setcounter{table}{0}
\setcounter{section}{0}
\renewcommand\thesection{\Alph{section}}
\renewcommand\thesubsection{\thesection.\arabic{subsection}}
\newpage

\section*{Supplementary Information}
\section{Machine learned interatomic potential: fitting and benchmarking}
\label{subsec:mlip}
A machine learned interatomic potential (MLIP) was trained to Rh-H interactions using the Atomic Cluster Expansion (ACE) formalism.\cite{drautz_atomic_2019} ACE was chosen for its speed and its ability to extrapolate well to structures outside of the domain of the training data. The MLIP was initially intended for the study of Au-Rh binary metal nanoparticles using training data that originated from a study on how the miscibility of AuRh nanoparticles can be modulated via surface passivation.\cite{chen_complete_2024} As such, Au interactions are included as well as Rh and H interactions, and the potential could be used in the future to simulate any combinations of these three elements. Guidance for accessing the potential for future use is included in the Data Availability section. 

\subsection{DFT training data}
The training data used to fit the MLIP was calculated with density functional theory (DFT) implemented in the Vienna Ab-initio Simulation Package (VASP).\cite{kresse_efficient_1996} The DFT calculations were performed with spin-polarization and a plane wave basis set, and the exchange and correlation energies were calculated using the Perdew-Burke-Ernzerhof (PBE) form of the generalized gradient approximation (GGA).\cite{perdew_generalized_1996} Non-periodic systems (i.e., surfaces and gases) were provided at least 10~Å of vacuum spacing to reduce self-interaction between periodic images. Calculations of bulk materials used $10\times 10\times 10$ k-point sampling, while calculations of surfaces and interfaces used $4\times 4\times 1$ k-point sampling, reducing the sampling along the direction perpendicular to the interface to 1 point, and calculations of isolated gaseous species used one k-point, i.e., the $\Gamma$ point. Gaussian smearing was applied with a width of 0.2 eV, a cutoff energy of 520 eV was applied for the plane wave basis set, and the electron–ion interactions were described by the projector augmented wave (PAW) method.\cite{kresse_ultrasoft_1999}

An initial set of 5455 structures was used to train the Au-Rh-H interatomic potential. This initial training set consisted of AuRh nanoparticles with sizes up to 314 atoms and different phase separations (alloy, heterodimer, core-shell) and structural defects (vacancies, icosahedral centers, etc.) as well as unary slabs, binary interfaces, and alloyed and unary bulk structures.

The model was subsequently refined with the incorporation of active learning structures, such that 13311 structures in total were used to fit the final potential. Active learning is performed by identifying gaps in the initial training dataset for which the potential energy surface is ill-defined. A metric called the extrapolation grade was used to help identify active learning structures.\cite{lysogorskiy_active_2023} For example, structures with the H$_{\text{top}}$ type hydrogen were added to the training set to confirm that the presence of this state was not an artifact of the potential. Additionally, \ce{H2} gas, small clusters of Au and Rh, H interstitials in bulk Au-Rh, and Au-Rh nanoparticles saturated with H adsorbates, were all added as part of active learning steps. 

\subsection{MLIP fitting}
The final potential architecture considers interaction distances of up to 7~Å and has 2-body, 3-body and 4-body interactions of all types (unary, binary, and ternary) with a total of 2502 basis functions and 3222 parameters. The $n_{\text{max}}$/$l_{\text{max}}$ for 2-body, 3-body and 4-body interactions are 10/0, 4/2, and 3/1 respectively. In order to help the potential fitting convergence, Au-Rh and H interactions were learned first by using training data that only had separated interactions, and then the fit was refined by updating the training data to include all interactions (including Au-H and Rh-H and Au-Rh-H interactions). Additionally, early stages of the fitting emphasized fitting the loss function to forces over energies ($\kappa$ = 0.9), with the final fit refining the energies ($\kappa$ = 0.02).

\begin{figure}[!htb]
    \centering
    \includegraphics[width=.5\linewidth]{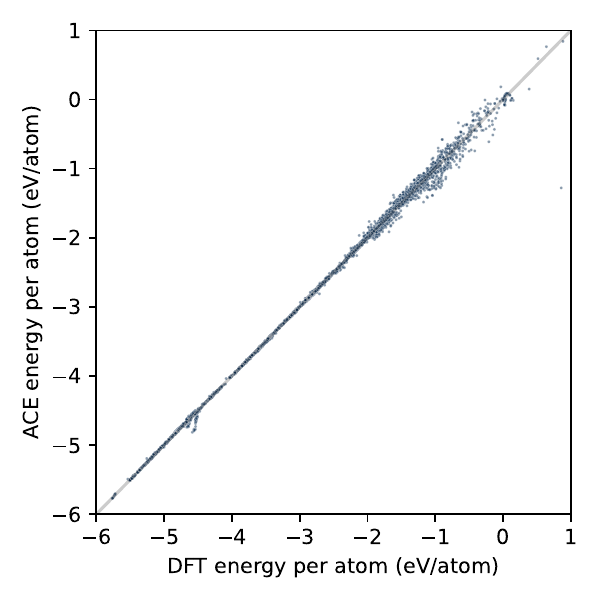}
    \caption{Parity plot of the ACE predicted and DFT calculated energies for the entire set of MLIP training data. The error is more pronounced for the very high energy structures that were added in active learning steps.}
    \label{fig:MLIPbenchmark}
\end{figure}

10 percent of the dataset was held out for testing, and the final root mean squared error (RMSE) on the test/train dataset was 33.30/38.37 meV/atom in energies and 145.33/153.53 meV/Å in forces. Looking at just the more realistic structures--those within 1 eV of the energy hull--the test/train metrics are 9.41/8.62 meV/atom in energies and 54.22/53.65 meV/Å in forces. Lower energy structures are much better represented by the potential, as can also be seen in the DFT/ACE energy parity plot given in Figure \ref{fig:MLIPbenchmark}. Considering only the initial dataset (binary nanoparticles, bulk, interfaces, and hydrogen adsorption to surfaces), the RMSE is 7.1 meV/atom. 

\section{Molecular dynamics simulations}
\label{subsec:simsetting}
\subsection{Simulation settings}
All simulations were carried out in LAMMPS with a Langevin thermostat held at 450~K and with a time step resolution of 0.5 fs.\cite{thompson_lammps_2022,lysogorskiy_performant_2021} Each simulation was pre-equilibrated by allowing gaseous hydrogen to adsorb to the metal surfaces, and then equilibrated further at the correct hydrogen concentration for at least 400 ps before commencing the production run. 200 ns of trajectory data was then collected for each system, where each hydrogen in the simulation contributes an individual trajectory. For example, a system with 100 hydrogen atoms would be simulated for 2 ns to achieve 200 ns of total trajectory data. Frames were printed every 10 fs, meaning that only processes with timescales slower than 10 fs can be resolved from this data. The largest simulation, which had 4939 atoms in total, could be simulated at a rate of 6.7 ns per day on a single CPU node with 64 processors. All simulations were rendered with the Ovito software.\cite{stukowski_visualization_2009}

\subsection{Thermostats}
The simulations were performed with a Langevin thermostat and a damping parameter setting of 1 ps, and other thermostats as well as a NVE simulation were tested to confirm that the choice of thermostat did not significantly bias the dynamical results (see Table \ref{tab:therm}). Each thermostat used the same geometry (5 nm nanoparticle with 1196 hydrogen atoms), simulation time (100 ns total), and MSM fitting parameters. Each tested thermostat, with the exception of a Langevin thermostat with a damping parameter of 10 ps, gave very similar timescale predictions for the association/dissociation reaction and for diffusion between facets.
\begin{table}
\begin{tabular}{lcc}
\multicolumn{1}{c}{Thermostat}                    & $\mathrm{H}_{(100)} \leftrightarrow \mathrm{H}_{(111)}$ & $\frac{1}{2}\mathrm{H}_{2(s)} \leftrightarrow \mathrm{H}_{(s)}$ \\ \hline
Langevin, damping parameter = 1 ps                & 119.4 ps                              & 58.6 ps                        \\
NVE                                               & 117.7 ps                              & 63.6 ps                       \\
Langevin, Rh atoms only, damping parameter = 1 ps & 122.6 ps                              & 58.7 ps                       \\
Langevin, damping parameter = 10 ps               & 136.3 ps                              & 114.5 ps                     
\end{tabular}
\caption{Benchmarking of timescale predictions using different thermostats for a test system of a 5 nm nanoparticle with 1196 hydrogen atoms.}
\label{tab:therm}
\end{table}

\section{MSM Fitting}
\label{subsec:msmfitting}

\subsection{Structure embedding}
The local environments of each hydrogen were represented with a subset of the ACE basis functions that comprised the Au-Rh-H MLIP. Interactions with Rh and H atoms up to 7Å away from each hydrogen were included, where the contributions from each interaction decayed to zero at this cutoff length. ACE basis functions for 2-body, 3-body and 4-body interactions were included with an $n_{\text{max}}$/$l_{\text{max}}$ of 4/0, 3/2, 2/1, respectively (131 total H-centered basis functions).

\subsection{Adsorption mode}
The adsorption mode was removed so as to focus the analysis entirely on the surface reactions. While the hydrogen adsorption/desorption mode is possible to be analyzed with the same methodology used here to investigate the hydrogen association/dissociation reaction, it would have required more data to be sufficiently sampled. Therefore, all trajectory frames for which the central hydrogen was classified as being in the gas phase was removed before analysis. As the simulations were sufficiently equilibrated, hydrogen atoms adsorbed and desorbed at an equal rate, and therefore the concentration of hydrogen atoms on the surface was relatively constant over each simulation.

\subsection{Hyperparameter selection}
It was essential to choose hyperparameters for the MSM construction that led to converged timescales over the entire range of simulations with different geometries and surface coverages. The hyperparameters that needed to be selected were the number of TICA dimensions, the number of Kmeans++ cluster centers, the lag time for the TICA projection fitting, and the lag time for the MSM model. The large nanoparticle, which has all of the states that are present in the small nanoparticle as well as the two slabs, was used as a test system for hyperparameter convergence tests.  

\begin{figure}[htbp]
    \centering
    \includegraphics[width=\linewidth]{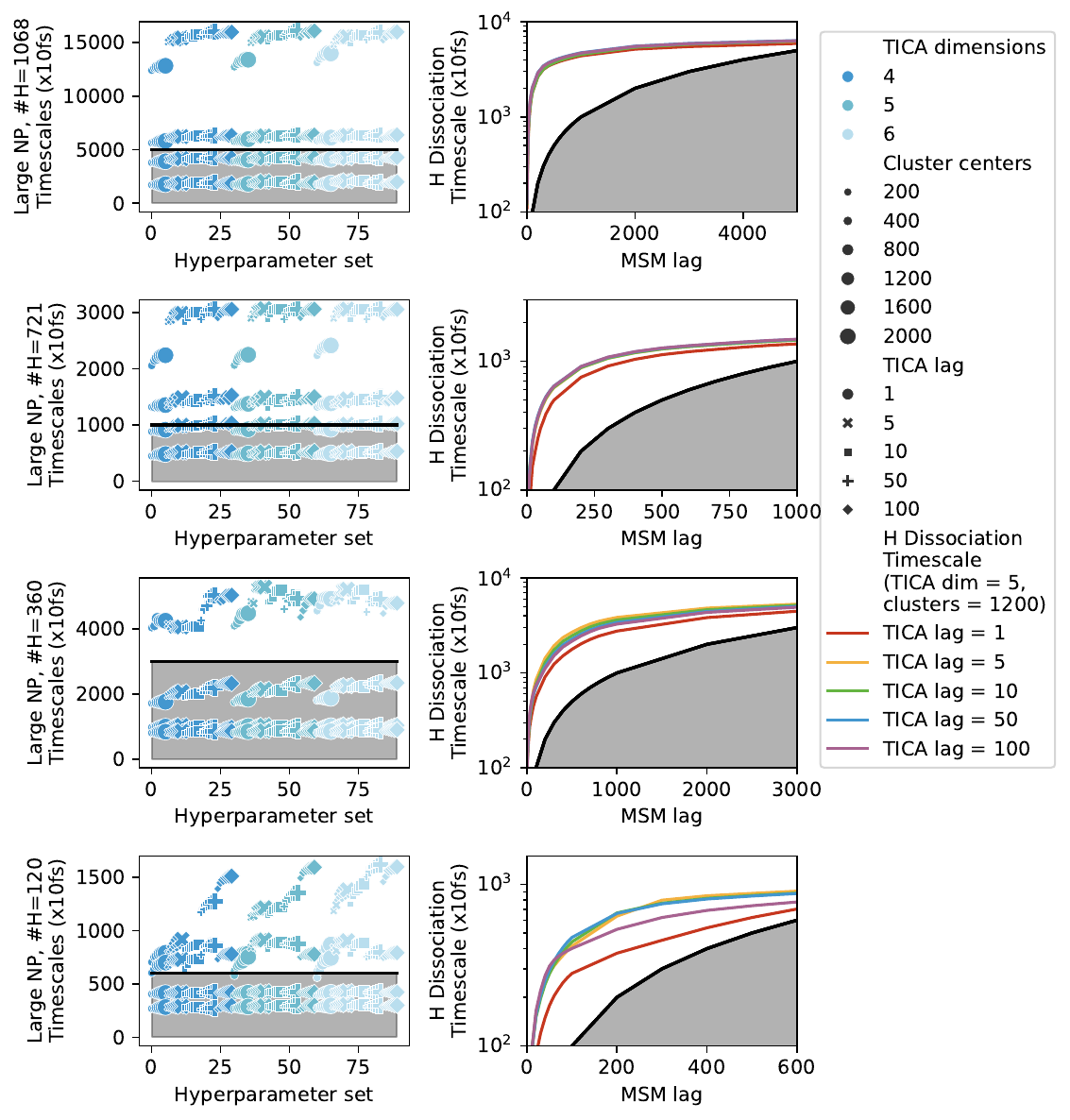}
    \caption{MSM hyperparameter convergence tests were performed for four of the simulations: large nanoparticles (5nm) with 120, 360, 721, and 1068 hydrogens. (Left) The number of TICA dimensions, the number of kmeans++ cluster centers, and the lag time for the TICA dimensionality reduction were all varied and tested for convergence in the first four slowest timescales. The MSM lag for these convergence tests are marked by the black line, with all timescales lower than this line  (Right) }
    \label{fig:hyperparameters}
\end{figure}

Implied timescales predicted by MSMs converge with a MSM time lag hyperparameter, as can be seen in Figure \ref{fig:flux}C. MSM lags were chosen to be 1/3 of the point in which the MSM lag would equal the association/dissociation timescale. A single MSM lag time would not be appropriate for all the simulations, because the implied timescales of interest were dramatically different for each simulation.

\subsection{Fluxes for representative systems}
In addition to the example simulation (large nanoparticle with 598 hydrogen atoms) that is discussed in detail in section 3.2, here we show a few other representative example simulations that have a range of different hydrogen coverages and rhodium geometries.

\begin{figure}[htbp]
    \centering
    \includegraphics[width=\linewidth]{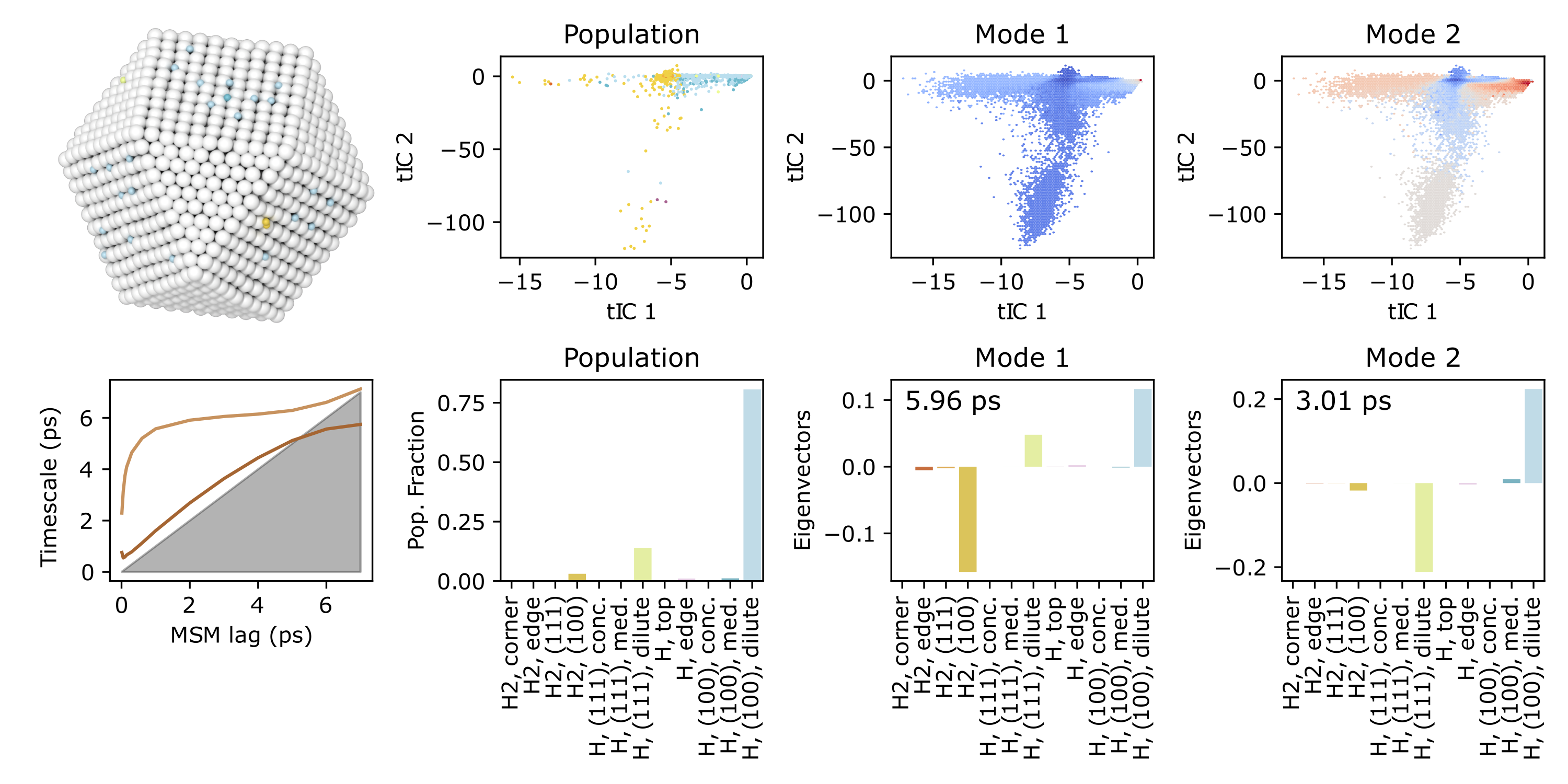}
    \caption{MSM analysis for a large (5 nm) nanoparticle with 60 hydrogen atoms (dilute coverage). Here, we see that at low concentrations of hydrogen, there is a near zero contribution to the H$\leftrightarrow\frac{1}{2}$\ce{H2} mode from \ce{H2} on corners and edges. Note that this simulation has poor convergence in the \ce{H_{(100)}}$\leftrightarrow$\ce{H_{(111)}} timescale.}
    \label{fig:fluxDilute}
\end{figure}

\begin{figure}[htbp]
    \centering
    \includegraphics[width=\linewidth]{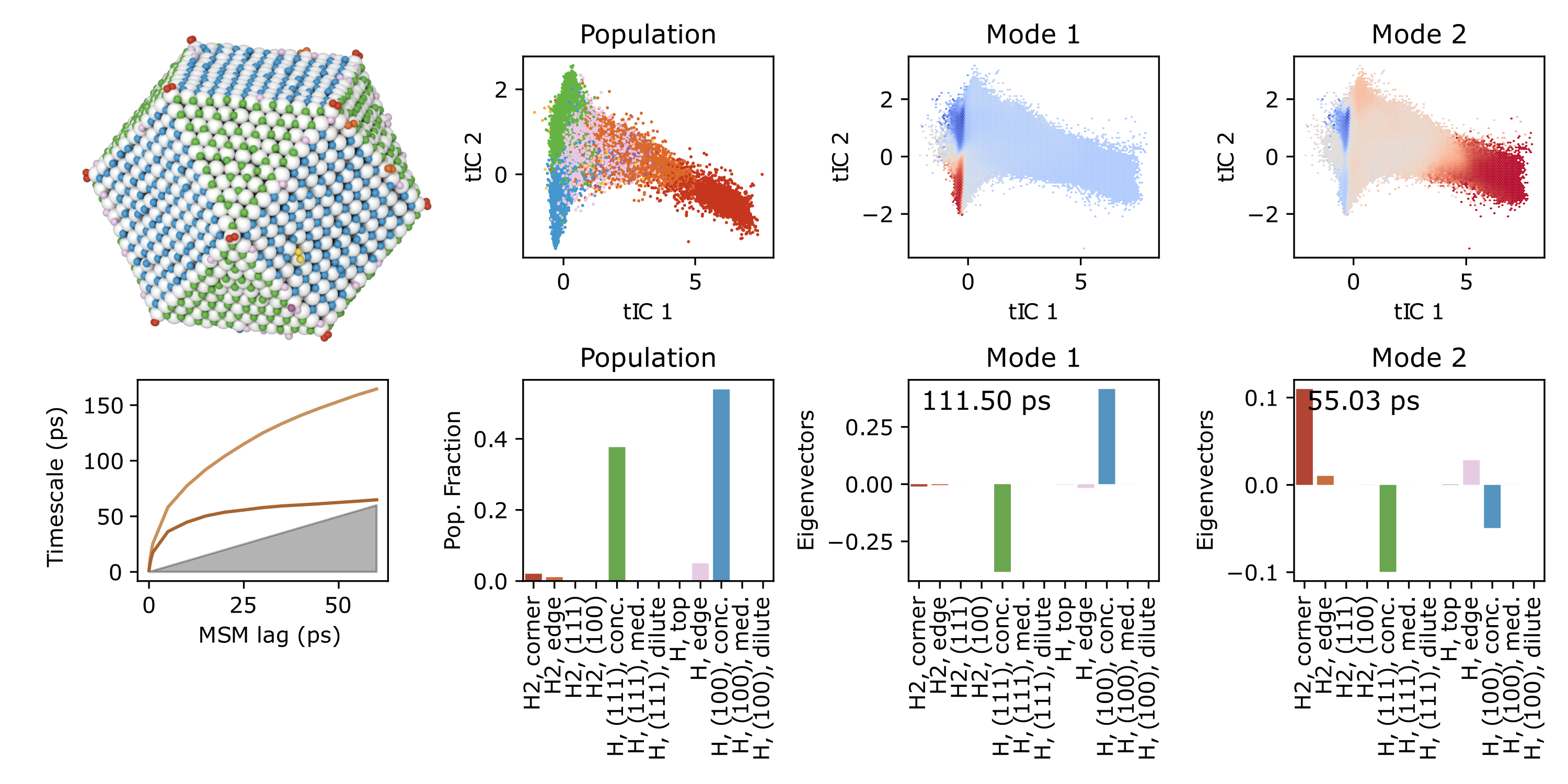}
    \caption{MSM analysis for a large (5 nm) nanoparticle with 1068 hydrogen atoms (high coverage). Here, the diffusion between facets is the slowest mode, followed by the H$\leftrightarrow\frac{1}{2}$\ce{H2} mode where the \ce{H2} states are mainly on the corners of the nanoparticle. Note that this simulation has poor convergence in the \ce{H_{(100)}}$\leftrightarrow$\ce{H_{(111)}} timescale for the chosen MSM lag.}
    \label{fig:fluxSaturated}
\end{figure}
\begin{figure}[htbp]
    \centering
    \includegraphics[width=\linewidth]{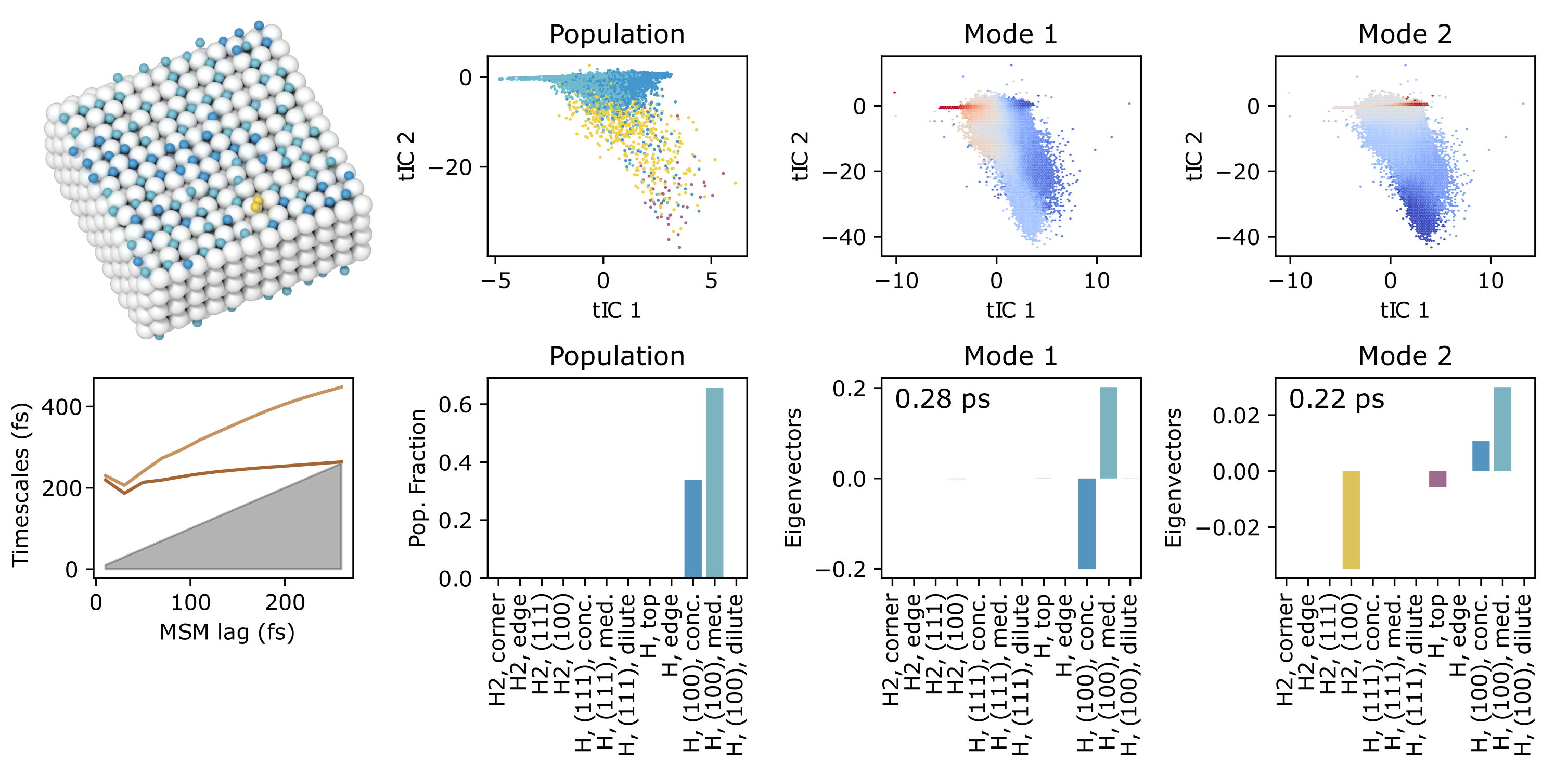}
    \caption{MSM analysis for a (100) slab surface with 215 hydrogen atoms (medium coverage). The slowest mode is the exchange between environments with many nearby hydrogen atoms and environments with fewer nearby hydrogen atoms (H$_{\text{(100), conc.}}\leftrightarrow$H$_{\text{(100), med.}}$). The next slowest mode is the exchange between \ce{H2} and H on the (100) facet. Note that this simulation has poor convergence in the \ce{H_{(100)}}$\leftrightarrow$\ce{H_{(111)}} timescale for the chosen MSM lag.}
    \label{fig:flux100}
\end{figure}
\begin{figure}[htbp]
    \centering
    \includegraphics[width=\linewidth]{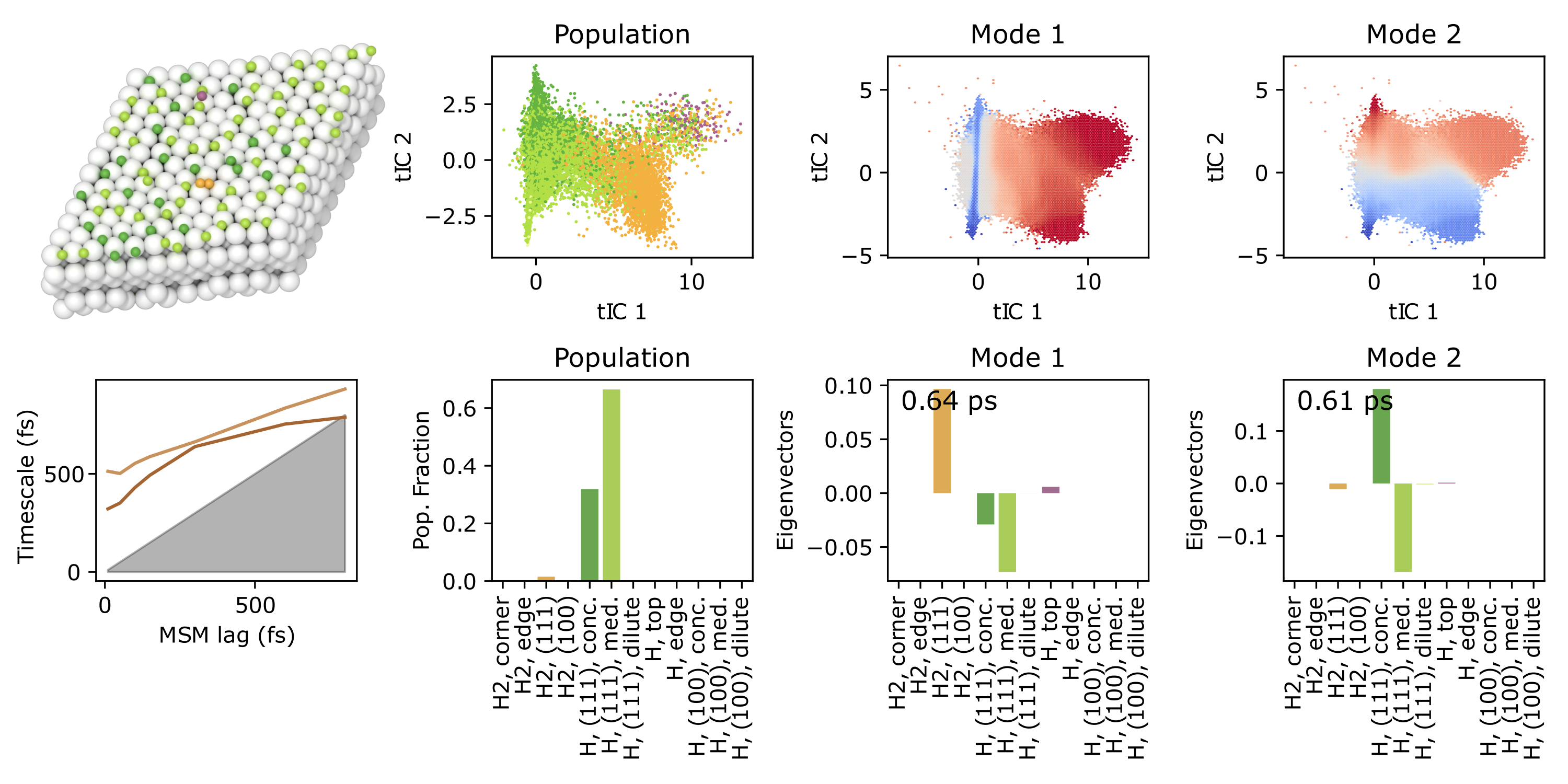}
    \caption{MSM analysis for a (111) slab surface with 165 hydrogen atoms (medium coverage). The slowest mode is the exchange between \ce{H2} and H on the (111) facet, and the next slowest mode is the exchange between environments with many nearby hydrogen atoms and environments with fewer nearby hydrogen atoms (H$_{\text{(111), conc.}}\leftrightarrow$H$_{\text{(111), med.}}$).}
    \label{fig:flux111}
\end{figure}

\begin{figure}[htbp]
    \centering
    \includegraphics[width=\linewidth]{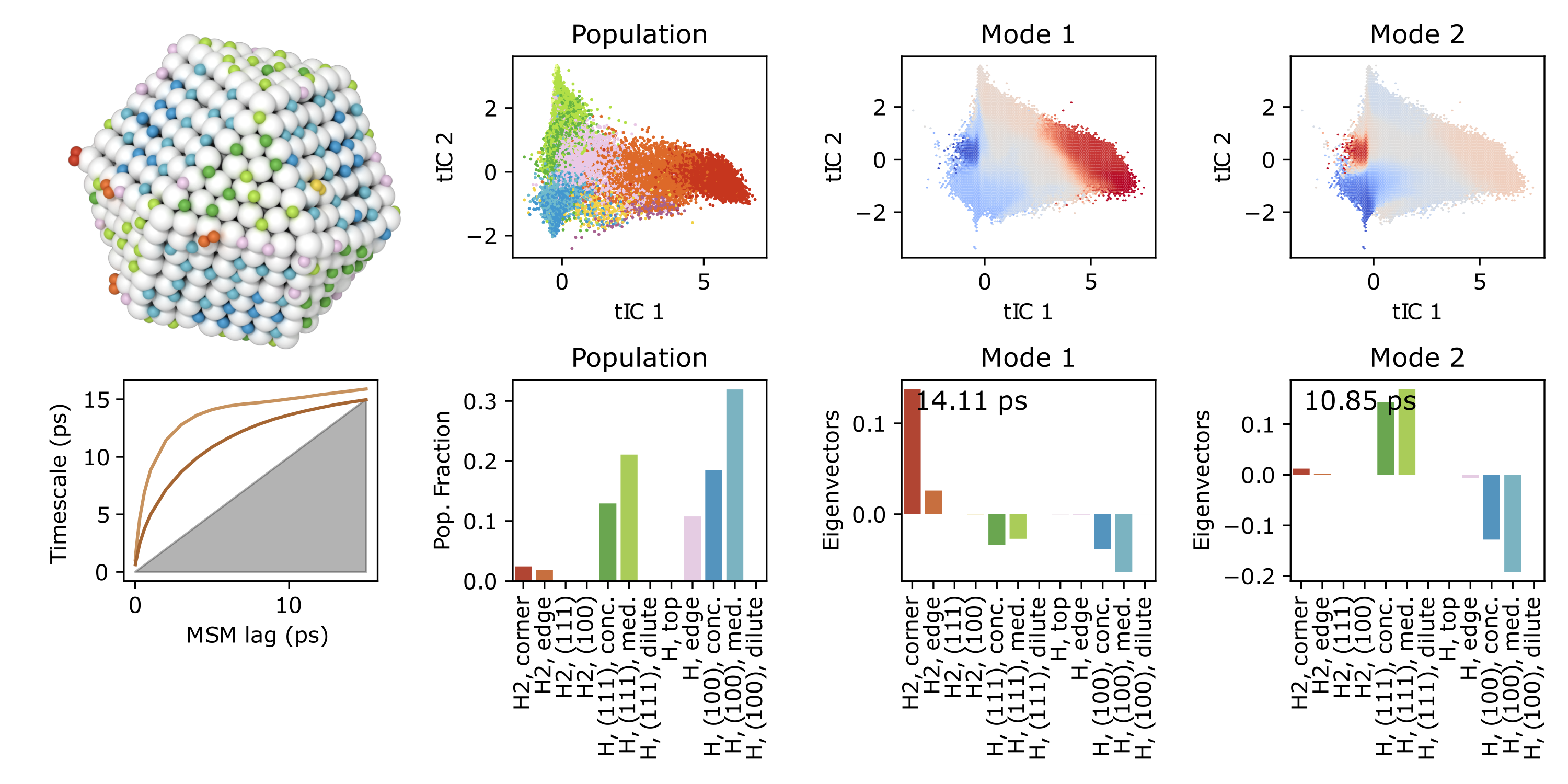}
    \caption{MSM analysis for a small nanoparticle (2 nm) with 266 hydrogen atoms (medium coverage). The slowest mode is the exchange between \ce{H2} on corners and edges and H on facets, and the next slowest mode is the exchange between H on each facet type (H$_{\text{(100)}}\leftrightarrow$H$_{\text{(111)}}$)}
    \label{fig:fluxSmallNP}
\end{figure}

\clearpage

\section{Committor and rate calculations}
\label{subsec:tpt}

Let $S_t^+$ be the first time after $t$ that the system is in $A$ or $B$, and $S_t^-$ be the last time before $t$ that the system is in $A$ or $B$.
Formally,
\begin{align}
    S_t^+ & = \min_{t'} \{ t' \ge t \mid \bx_{t'} \in A \cup B \} \quad , \\
    S_t^- & = \max_{t'} \{ t' \le t \mid \bx_{t'} \in A \cup B \} \quad .
\end{align}
The ``stopped'' transition operator that appears in \eqref{eq:committor} is defined to be
\begin{equation}
    \mathcal{S}(\tau) f(\bx) = \mathbbm{E}[f(\bx_{\min(t+\tau,S_t^+)}) \mid \bx_t = \bx] \quad .
\end{equation}

To estimate the committor using dynamical Galerkin approximation, \cite{thiede_galerkin_2019,strahan_long-time-scale_2021} we approximate the committor by the basis expansion
\begin{equation}
    q^+(\bx) = \sum_i c_i \mathbbm{1}_i(\bx) + \mathbbm{1}_B(\bx) \quad ,
\end{equation}
where the sum is over intermediate microstates, $\mathbbm{1}_i$ is the indicator function on intermediate microstate $i$, $\mathbbm{1}_B$ is the indicator function on the product state $B$, and $\bf{c}$ is a vector of unknown coefficients.
We solve for $\bf{c}$ from the weak form of \eqref{eq:committor},
\begin{equation}
    \mathbbm{E}[\mathbbm{1}_i(\bx) q^+(\bx)] = \mathbbm{E}[\mathbbm{1}_i(\bx) \mathcal{S}(\tau) q^+(\bx)] \quad .
\end{equation}
For a trajectory $\bx^T$, assuming detailed balance, we evaluate
\begin{align}
    L_{ij}
    & = \mathbbm{E}[\mathbbm{1}_i (\mathcal{S}(\tau) \mathbbm{1}_j - \mathbbm{1}_j)] \\
    \label{eq:dga_L}
    & \begin{aligned}[b]
    {} = \frac{1}{2(T-\tau)} \sum_{t=1}^{T-\tau} \biggl(
        & \mathbbm{1}_i(\bx_t) \left(\mathbbm{1}_j(\bx_{\min(t+\tau,S_t^+)}) - \mathbbm{1}_j(\bx_t)\right) \\
        & + \mathbbm{1}_i(\bx_{t+\tau}) \left(\mathbbm{1}_j(\bx_{\max(t,S_{t+\tau}^-)}) - \mathbbm{1}_j(\bx_{t+\tau})\right)
    \biggr) \quad ,
    \end{aligned}
    \\
    r_i
    & = \mathbbm{E}[\mathbbm{1}_i (\mathcal{S}(\tau) \mathbbm{1}_B - \mathbbm{1}_B)] \\
    \label{eq:dga_r}
    & \begin{aligned}[b]
    {} = \frac{1}{2(T-\tau)} \sum_{t=1}^{T-\tau} \biggl(
        & \mathbbm{1}_i(\bx_t) \left(\mathbbm{1}_B(\bx_{\min(t+\tau,S_t^+)}) - \mathbbm{1}_B(\bx_t)\right) \\
        & + \mathbbm{1}_i(\bx_{t+\tau}) \left(\mathbbm{1}_B(\bx_{\max(t,S_{t+\tau}^-)}) - \mathbbm{1}_B(\bx_{t+\tau})\right)
    \end{aligned}
    \biggr) \quad .
\end{align}
In \eqref{eq:dga_L} and \eqref{eq:dga_r}, the expectation is approximated by an average over length $\tau$ rolling windows of the trajectory $\bx^T$ (first term of the summand) and the time-reversed trajectory (second term of the summand).
To obtain $\bf{c}$, we solve the linear system
\begin{equation}
    \bf{L} \bf{c} = -\bf{r} \quad .
\end{equation}

As in \eqref{eq:dga_L} and \eqref{eq:dga_r}, we estimate the expression for the net reactive flux, \eqref{eq:rate}, \cite{strahan_long-time-scale_2021,lorpaiboon_exact_2026} by approximating the expectation by an average over rolling windows.
For a single trajectory $\bx^T$, we evaluate \eqref{eq:rate} using the estimated committor as
\begin{align}
    F_{AB} = \frac{1}{2 (T-\tau)} \sum_{t=1}^{T-\tau} \Biggl(
        & \frac{1}{\tau} \sum_{t'=t}^{t+\tau-1} q^-(\bx_{\max(t,S_{t'}^-)}) \left(q^+(\bx_{t'+1}) - q^+(\bx_{t'})\right) q^+(\bx_{\min(t+\tau,S_{t'+1}^+)})
        \nonumber \\
        & + \frac{1}{\tau} \sum_{t'=t}^{t+\tau-1} q^+(\bx_{\max(t,S_{t'}^-)}) \left(q^-(\bx_{t'+1}) - q^-(\bx_{t'})\right) q^-(\bx_{\min(t+\tau,S_{t'+1}^+)})
    \Biggr) \quad .
\end{align}

\section{NEB calculations}
\label{subsec:neb}
\begin{figure}[htbp]
    \centering
    \includegraphics[width=\linewidth]{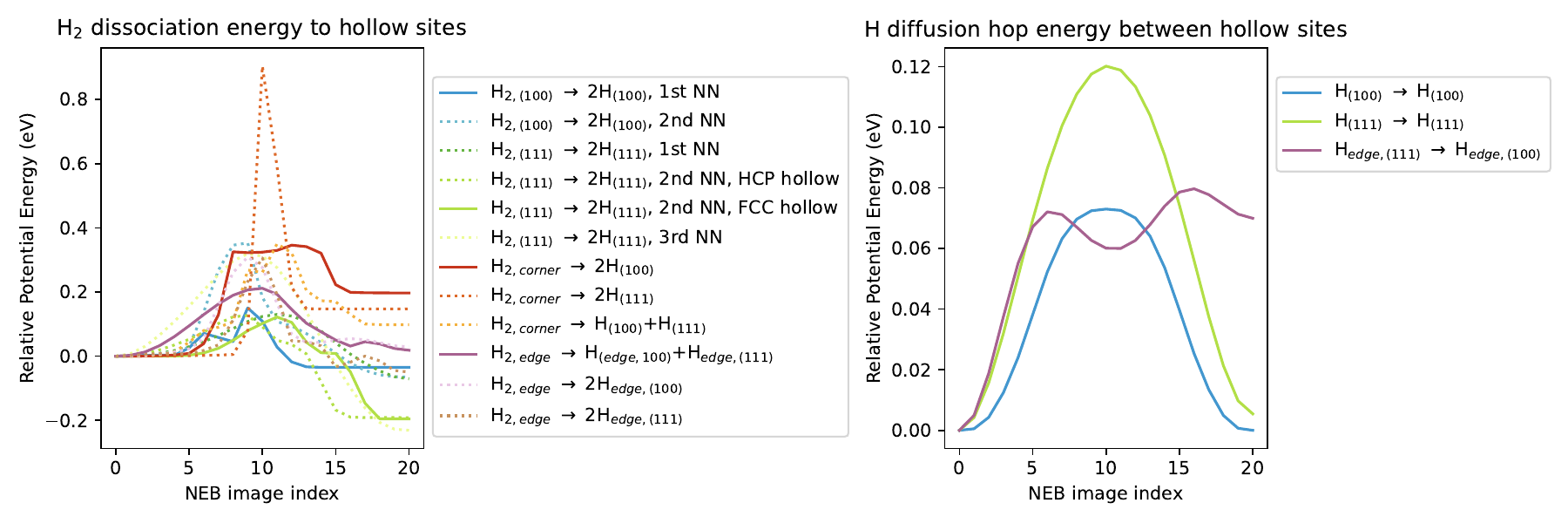}
    \caption{NEB calculations for energy barrier estimates of hydrogen association, dissociation, and diffusion.}
    \label{fig:NEB}
\end{figure}

Standard TST (transition state theory) calculations for the H $\rightarrow$ \ce{H2} association and \ce{H2}$\rightarrow$ H dissociation rates were calculated using energy barriers generated fom nudged elastic band (NEB) calculations (Figure \ref{fig:NEB}).~\cite{jonsson_nudged_1998} The NEB calculations were performed with the climbing image algorithm implemented in ASE.~\cite{henkelman_climbing_2000} \ce{H2} molecules were initialized atop rhodium atoms on the facets, corners and edges of nanoparticles, and were split into neighboring hollow sites. The initial and final states were geometrically relaxed prior to performing the NEB calculation. 20 images, initialized as a linear interpolation between the initial and final images, were used. The TST rates were then calculated with the following approximation for the rate constant:
\\
\begin{equation}
k = k_b \frac{T}{h}e^{-\Delta G_t/(k_bT)}
\end{equation}
\\
Note that the energy barriers, $\Delta G_t$, were not vibrationally corrected. Diffusion barriers were also calculated to confirm that the diffusion energy barriers were lower than that of association and dissociation.

\end{document}